\documentclass[acmsmall,screen,table]{acmart}
\makeatletter
\AddToHook{cmd/added/before}{\color{black}}
\makeatother
\usepackage{amsmath,amsfonts}
\usepackage{algorithmic}
\usepackage{graphicx}
\usepackage{textcomp}
\usepackage{xcolor}
\def\BibTeX{{\rm B\kern-.05em{\sc i\kern-.025em b}\kern-.08em
    T\kern-.1667em\lower.7ex\hbox{E}\kern-.125emX}}
\usepackage{listings}
\usepackage[utf8]{inputenc}
\usepackage{color}
\usepackage{tabularray}
\usepackage{graphicx}
\usepackage{colortbl}
\usepackage{threeparttable}
\usepackage{multirow}
\usepackage{fancybox}
\usepackage{setspace}
\usepackage{caption}
\usepackage{subcaption}
\usepackage{algorithm}
\usepackage{algorithmic}
\usepackage[most]{tcolorbox}
\usepackage{adjustbox}
\usepackage{tabulary}
\usepackage{nicematrix}
\usepackage{threeparttable}
\usepackage{hyperref}
\usepackage{balance}
\usepackage{enumitem}
\usepackage{booktabs}
\usepackage{xspace}
\usepackage{collab}
\usepackage{verbatim}

\usepackage[commandnameprefix=ifneeded]{changes} 
\usepackage{bbding}
\usepackage{makecell} 
\usepackage{fontawesome5} 
\newcommand{\fixedwidth}[1]{\texttt{#1}}

\usepackage{multirow}      
\usepackage{booktabs}      
\usepackage{tikz}           

\newcommand{\heatcell}[2]{\cellcolor{maxcolor!#1!mincolor}#2}

\definecolor{mincolor}{RGB}{223, 246, 246} 
\definecolor{maxcolor}{RGB}{113, 195, 185}

\definecolor{codegreen}{rgb}{0,0.6,0}
\definecolor{codegray}{rgb}{0.5,0.5,0.5}
\definecolor{codepurple}{rgb}{0.58,0,0.82}
\definecolor{backcolour}{rgb}{0.95,0.95,0.92}
\definecolor{cerulean}{rgb}{0.0, 0.48, 0.65}
\definecolor{ceruleanblue}{rgb}{0.16, 0.32, 0.75}
\definecolor{cadmiumred}{rgb}{0.89, 0.0, 0.13}
\definecolor{grey}{rgb}{0.9, 0.9, 0.9}
\definecolor{viol}{RGB}{134,0,175}
\definecolor{githubgreen}{RGB}{204, 255, 204}
\definecolor{githubred}{RGB}{255, 224, 224}
\definecolor{mygray}{rgb}{0.8,0.8,0.8}
\definecolor{lightyellow}{rgb}{1,1,0.8}
\lstdefinestyle{test}{
    language={sh},
    moredelim=**[is][\color{red}]{~}{~},
    basicstyle=\ttfamily, 
}

\definecolor{ballblue}{rgb}{0.13, 0.67, 0.8}
\collabAuthor{gb}{ballblue}{Guangba Yu}
\collabAuthor{ry}{red}{Renyi Zhong}
\collabAuthor{jx}{green}{Jinxi Kuang}
\collabAuthor{yc}{blue}{Yichen Li}
\collabAuthor{yt}{orange}{Yintong Huo}
\collabAuthor{ww}{violet}{Wenwei Gu}

\AtBeginDocument{%
  \providecommand\BibTeX{{%
    Bib\TeX}}}

\setcopyright{acmlicensed}
\copyrightyear{2018}
\acmYear{2018}
\acmDOI{XXXXXXX.XXXXXXX}

\acmConference[Conference acronym 'XX]{Make sure to enter the correct
  conference title from your rights confirmation emai}{June 03--05,
  2018}{Woodstock, NY}
\acmISBN{978-1-4503-XXXX-X/18/06}

\begin{document}

\title{Larger Is Not Always Better: Exploring Small Open-source Language Models in Logging Statement Generation}

\author{Renyi Zhong}
\email{ryzhong22@cse.cuhk.edu.hk}
\affiliation{%
  \institution{The Chinese University of Hong Kong}
  \country{Hong Kong}
}
\author{Yichen Li}
\email{ycli21@cse.cuhk.edu.hk}
\affiliation{%
  \institution{The Chinese University of Hong Kong}
  \country{Hong Kong}
}

\author{Guangba Yu}
\email{gbyu@cse.cuhk.edu.hk}
\affiliation{%
  \institution{The Chinese University of Hong Kong}
  \country{Hong Kong}
}

\author{Wenwei Gu}
\email{wwgu@cse.cuhk.edu.hk}
\affiliation{%
  \institution{The Chinese University of Hong Kong}
  \country{Hong Kong}
}

\author{Jinxi Kuang}
\email{jxkuang22@cse.cuhk.edu.hk}
\affiliation{%
  \institution{The Chinese University of Hong Kong}
  \country{Hong Kong}
}

\author{Yintong Huo}
\email{ythuo@smu.edu.sg}
\affiliation{%
  \institution{Singapore Management University}
  \country{Singapore}
}

\author{Michael R. Lyu}
\email{lyu@cse.cuhk.edu.hk}
\affiliation{%
  \institution{The Chinese University of Hong Kong}
  \country{Hong Kong}
}

\begin{abstract}
Developers use logging statements to create logs that document system behavior and aid in software maintenance. As such, high-quality logging is essential for effective maintenance; however, manual logging often leads to errors and inconsistency. Recent methods emphasize using large language models (LLMs) for automated logging statement generation, but these present privacy and resource issues, hindering their suitability for enterprise use.
This paper presents the first large-scale empirical study evaluating small open-source language models (SOLMs) for automated logging statement generation. We evaluate four prominent SOLMs using various prompt strategies and parameter-efficient fine-tuning techniques, such as Low-Rank Adaptation (LoRA) and Retrieval-Augmented Generation (RAG). Our results show that fine-tuned SOLMs with LoRA and RAG prompts, particularly Qwen2.5-coder-14B, outperform existing tools and LLM baselines (e.g., Claude3.7 sonnet and GPT4o) in predicting logging locations and generating high-quality statements, with robust generalization across diverse repositories. These findings highlight SOLMs as a privacy-preserving, efficient alternative for automated logging.
\end{abstract}


\maketitle

\section{Introduction}

Logs are textual records generated during software execution to capture runtime events, states, and contextual information~\cite{zhong2025towards}. A typical log statement consists of three components: a verbosity level, logging variables, and logging texts~\cite{he2021survey}.
In particular, as the example shown below, the logging level (e.g., \texttt{debug}) reflects the event's severity; the logging variables (e.g., \texttt{terminalState}) hold critical run-time data about system states; meanwhile, the logging text (e.g., \texttt{Stopping the checkpoint services with state}) offers a static explanation of the system's actions.
\begin{center}
    \fbox
    {\shortstack[l]{
    \fixedwidth{log.debug("Stopping the checkpoint services with state \{ \}.", terminalState); }
    }}
\end{center}

High-quality logs provide actionable insights for developers to diagnose failures, optimize system behavior, and ensure reliability. However, the absence or inadequacy of logging statements can severely hinder downstream tasks such as anomaly detection~\cite{zhangDeepTraLogTracelogCombined2022} and failure diagnosis~\cite{jiang2025l4,huang2024demystifying}, leading to prolonged debugging cycles and increased maintenance costs. Consequently, the strategic placement and content of log statements directly influence the effectiveness of software maintenance and evolution~\cite{soldani2022anomaly,gu2025kpiroot}. 

Despite their critical role in software maintenance, producing high-quality logs manually is far from straightforward for developers. First, the absence of universal logging guidelines leads to inconsistent practices, where log quality, granularity, and utility vary widely based on individual developers’ expertise. This inconsistency complicates log analysis and reduces their diagnostic value~\cite{chen2019extracting}. Second, developers face the difficult task of balancing log quantity and quality: over-logging burdens systems with excessive data, while under-logging risks missing critical information~\cite{yaoLog4PerfSuggestingLogging2018}. Third, the cognitive and time-intensive nature of manual logging further exacerbates these issues, as developers must anticipate complex system behaviors and failure points, often resulting in logs that are either too vague or overly detailed~\cite{hassani2018studying}. Additionally, maintaining log relevance over time is challenging, as software evolution can render existing logs obsolete~\cite{zhong2025LogUpdater}. These challenges highlight the need for automated logging statement generation (hereafter referred to simply as `automated logging') solutions that can consistently produce high-quality logs.

To address these challenges, at the early stage, researchers have explored automated logging techniques by decomposing the problem into sub-tasks, such as where-to-log (identifying code locations for logging)~\cite{zhaoLog20FullyAutomated2017,liWhereShallWe2020}, what-to-log (generating static content and dynamic variables)~\cite{dingLoGenTextAutomaticallyGenerating2022,liuWhichVariablesShould2019}, and log-level suggestion~\cite{heng2024studying,liDeepLVSuggestingLog2021,liuTeLLLogLevel2022}. However, these fragmented approaches lack integration into a unified, end-to-end framework for generating complete logging statements.
Recent advances in pre-trained language models (LMs) have opened new avenues for automated logging. LANCE~\cite{mastropaoloUsingDeepLearning2022} and LEONID~\cite{mastropaolo2024log} pioneered the use of sequence-to-sequence models (e.g., T5~\cite{raffel2020exploring}) to generate logging statements directly from code contexts. Subsequent tools, such as Fastlog~\cite{xie2024fastlog}, Unilog~\cite{xu2024unilog}, and SCLogger~\cite{li2024go}, further leveraged the large language model (LLM) to improve logging quality. In particular, Unilog and SCLogger adopted prompt-based methods with LLMs such as GPT-3.5 and Codex, achieving state-of-the-art performance by harnessing the code comprehension and natural language generation capabilities of LLMs. 

Despite their effectiveness, LLM-based logging tools~\cite{xie2024fastlog,xu2024unilog,li2024go} face several limitations in enterprise settings. 
\begin{itemize}[leftmargin=*]
    \item \textbf{Privacy Risks.}  Sending proprietary code to commercial LLM APIs, such as OpenAI’s, risks exposing sensitive intellectual property~\cite{yao2024survey}. For instance, Samsung banned employee use of ChatGPT and other generative AI tools after an engineer accidentally leaked sensitive source code to ChatGPT~\cite{kharpal2023samsung}.
    \item \textbf{Style Misalignment.} LLMs, trained on generic datasets, struggle to generate logs that align with enterprise-specific logging styles, such as unique verbosity levels or error prioritization requirements, limiting their utility for organizational needs~\cite{he2022empirical,gu2022logging}. 
\end{itemize}
To address these issues, enterprises often consider fine-tuning and deploying open-source LLMs in private environments. However, this approach demands substantial computational resources, including thousands of GPU hours, which is impractical for many resource-constrained organizations~\cite{chien2023reducing,imani2024context}. These limitations necessitate more accessible and efficient solutions for automated logging.

Small open-source language models (SOLMs), defined as open-source models with fewer than 14 billion parameters, have gained traction in software engineering tasks, such as program repair~\cite{silva2023repairllama} and comment rectification~\cite{rong2024code}, making them a promising solution for automated logging. By deploying SOLMs locally on consumer-grade hardware, such as an A100 GPU, enterprises can safeguard proprietary code, eliminating privacy risks associated with commercial APIs~\cite{yang2024chain}. Moreover, SOLMs require significantly fewer computational resources, reducing costs and aligning with sustainable computing goals~\cite{imani2024context}. Additionally, SOLMs can be fine-tuned on enterprise-specific codebases to produce logs that meet unique organizational requirements, such as specific formats or compliance standards~\cite{lu2023llama,rong2024code}. For resource-constrained enterprises, SOLMs offer a practical balance of privacy, efficiency, and adaptability, making their application to automated logging highly attractive. However, to the best of our knowledge, no studies have systematically explored the effectiveness of SOLMs in automated logging.


To fill this significant gap,  in this paper, we conduct an empirical study on four prominent SOLMs, namely LLaMA~\cite{grattafiori2024llama}, Mistral~\cite{jiang2023mistral7b}, CodeLLaMA~\cite{roziere2023code}, and Qwen2.5coder~\cite{hui2024qwen2}, to explore the potential utility of SOLMs in automatic generation of logging statements. We pose four research questions to comprehensively assess the potential of SOLMs.

\textbf{RQ1: What are the most effective prompting strategies for using SOLMs in logging generation?} Different prompting techniques (e.g., in-context learning (ICL)~\cite{brown2020language}, chain-of-thought (COT)~\cite{wei2022chain}, retrieval-augmented generation (RAG)~\cite{lewis2020retrieval}) can influence the performance of SOLMs without the need for retraining. Gaining insight into their effects can help improve the generation of logging statements across different contexts. 

\textbf{\textit{Result.}} RAG outperforms ICL and COT, significantly enhancing the performance of logging automation task.

\textbf{RQ2: What is the best tuning strategy using SOLMs for automated logging?} Many strategies such as parameter-efficient fine-tuning (PEFT) techniques~\cite{houlsby2019parameter,hu2022lora}, model size, and model type may influence the efficacy. Thus, we further investigate the extent of their impact, which may offer insights into the optimal selection of strategies for enhancing SOLM performance.

\textbf{\textit{Result.}} LoRA~\cite{hu2022lora} demonstrates the most consistent and superior results when fine-tuning with PEFT techniques. For models with more than 3B parameters, performance in generating logging statements improves with more parameters, but the increased computational costs indicate a trade-off between performance and resource costs. The instruct variant of the SOLM model outperforms its base counterpart, benefiting from its instruction-tuned foundation.

\textbf{RQ3: How effectively do SOLMs compare to existing methods and LLM baselines in automated logging?} Upon recognizing the optimal strategies for employing SOLMs, we aim to investigate the performance of SOLMs in automated logging compared to existing methods and the prominent LLMs.

\textbf{\textit{Result.}} Fine-tuned SOLMs, particularly Qwen2.5-coder-14B, outperform both existing methods and LLMs across all evaluated metrics, demonstrating superior logging location accuracy and statement quality. The result of the LLM-based judger further supports the high quality of SOLM-generated statements.

\added{\textbf{RQ4: How robust are SOLMs in handling diverse real-world scenarios for logging statement generation?} We assess robustness by evaluating their generalization across diverse repositories and analyzing the performance impact of varying code lengths.}

\added{\textbf{\textit{Result.}} SOLMs demonstrate strong robustness, generalizing effectively to unseen repositories. Performance varies with code length, revealing a trade-off between positioning accuracy and content quality. We also find that similar logging practices across projects significantly boost generalization.}


\begin{table}[tbp]
\centering
\small
\caption{\added{Summarization of Key Findings and Implications in This Paper.}}
\label{tab:key-summarization}
\resizebox{\textwidth}{!}{%
\begin{tabular}{l||l}
\toprule
    \textbf{Key Findings}   &   \textbf{Key Implications \& Actionable Advice} \\
\toprule
$>>$ RQ1 \& RQ2: Prompting and Tuning Strategies & \\
\makecell[l]{\faHandPointRight[regular]~ Without fine-tuning, general-purpose models outperform code-specific \\ ones due to better instruction-following capabilities. (Finding 1)} & \multirow{5}{*}{\makecell[l]{\faArrowAltCircleRight[regular] An effective optimization blueprint for SOLMs is identified: \\ use instruct-tuned models as a base, apply LoRA for task \\ specialization, and integrate RAG for instance-specific context. \\ This holistic methodology offers a practical path for building \\ specialized logging tools that are not only resource-efficient \\ but also capable of achieving state-of-the-art performance.}} \\
\makecell[l]{\faHandPointRight[regular]~ Retrieval-Augmented Generation (RAG) is the most effective \\ prompting strategy for un-tuned SOLMs. (Finding 2)} & \\
\makecell[l]{\faHandPointRight[regular]~ LoRA is the most effective PEFT technique; instruct-tuned models \\ are the best foundation for fine-tuning. (Findings 3, 5)} & \\
\makecell[l]{\faHandPointRight[regular]~ A performance-resource sweet spot exists for models with 3B+ \\ parameters; a single fine-tuning epoch is often sufficient. (Findings 4, 6)} & \\
\makecell[l]{\faHandPointRight[regular]~ The combination of LoRA and RAG is synergistic and yields \\ optimal performance. (Finding 7)} & \\
\midrule
$>>$ RQ3: Baseline Comparison & \\
\makecell[l]{\faHandPointRight[regular]~ Fine-tuned SOLMs outperform existing methods and larger LLM \\ baselines across all evaluation metrics. (Finding 8)} & \makecell[l]{\faArrowAltCircleRight[regular] The "larger is better" paradigm is challenged. For specialized \\ logging tasks, a well-optimized SOLM is a more effective and \\ practical alternative to large, proprietary LLMs.} \\
\midrule
$>>$ RQ4: Robustness and Generalization & \\
\makecell[l]{\faHandPointRight[regular]~ SOLMs show strong generalization, which is enhanced by training \\ on data with consistent logging practices. (Findings 9, 10)} & \multirow{3}{*}{\makecell[l]{\faArrowAltCircleRight[regular] SOLMs are robust enough for real-world deployment, but this \\ requires strategic considerations. Data curation focused on \\ consistent standards is key to generalization. Tool design \\ should also account for known model behaviors, like the \\ performance trade-off with code length.}} \\
\makecell[l]{\faHandPointRight[regular]~ Code length introduces a trade-off: positioning accuracy decreases \\ while content generation quality improves. (Findings 11, 13)} & \\
\makecell[l]{\faHandPointRight[regular]~ The LoRA+RAG strategy remains the most robust choice across \\ different code lengths and complexities. (Finding 12)} & \\
\bottomrule
\end{tabular}
}
\end{table}

\added{In Table~\ref{tab:key-summarization}, we summarize the thirteen findings and three implications with actionable advice.}

\textbf{Contributions.} To sum up, in this paper, we make the following contributions:

\begin{itemize}
    \item We conduct the first large-scale empirical study assessing the effectiveness of SOLMs for automated logging. Our findings demonstrate their capability to produce contextually accurate and syntactically correct logging statements that rival or exceed the performance of existing specialized methods and LLMs.
    \item We investigate and identify effective strategies for optimizing SOLM performance in the context of automated logging. This includes demonstrating that RAG significantly enhances performance and that Low-Rank Adaptation stands out as a highly effective PEFT technique.
    \item We showcase the practical advantages of fine-tuned SOLMs, including their robust generalization capabilities across diverse and previously unseen code repositories. This research highlights that SOLMs can maintain high performance in varied settings and offer benefits such as local deployment for data privacy and alignment with enterprise-specific logging styles.
    \item To facilitate further research and encourage practical adoption in the field of automated logging, we publicly release our source code, datasets, and comprehensive experimental results~\cite{artifacts}.
\end{itemize}

%

\textbf{Paper Organizations.} Section 2 discusses the background. Section 3 describes the experimental design of our study. Section 4 presents the experimental results. Section 5 introduces a case study, some implications, the advantages of using SOLMs for automated logging, and potential future work directions. Section 6 discusses threats to validity. Section 7 introduces the related work. Section 8 concludes the paper.

\section{Background}

\subsection{Problem Definition}

\begin{figure}
    \centering
    \includegraphics[width=0.7\linewidth]{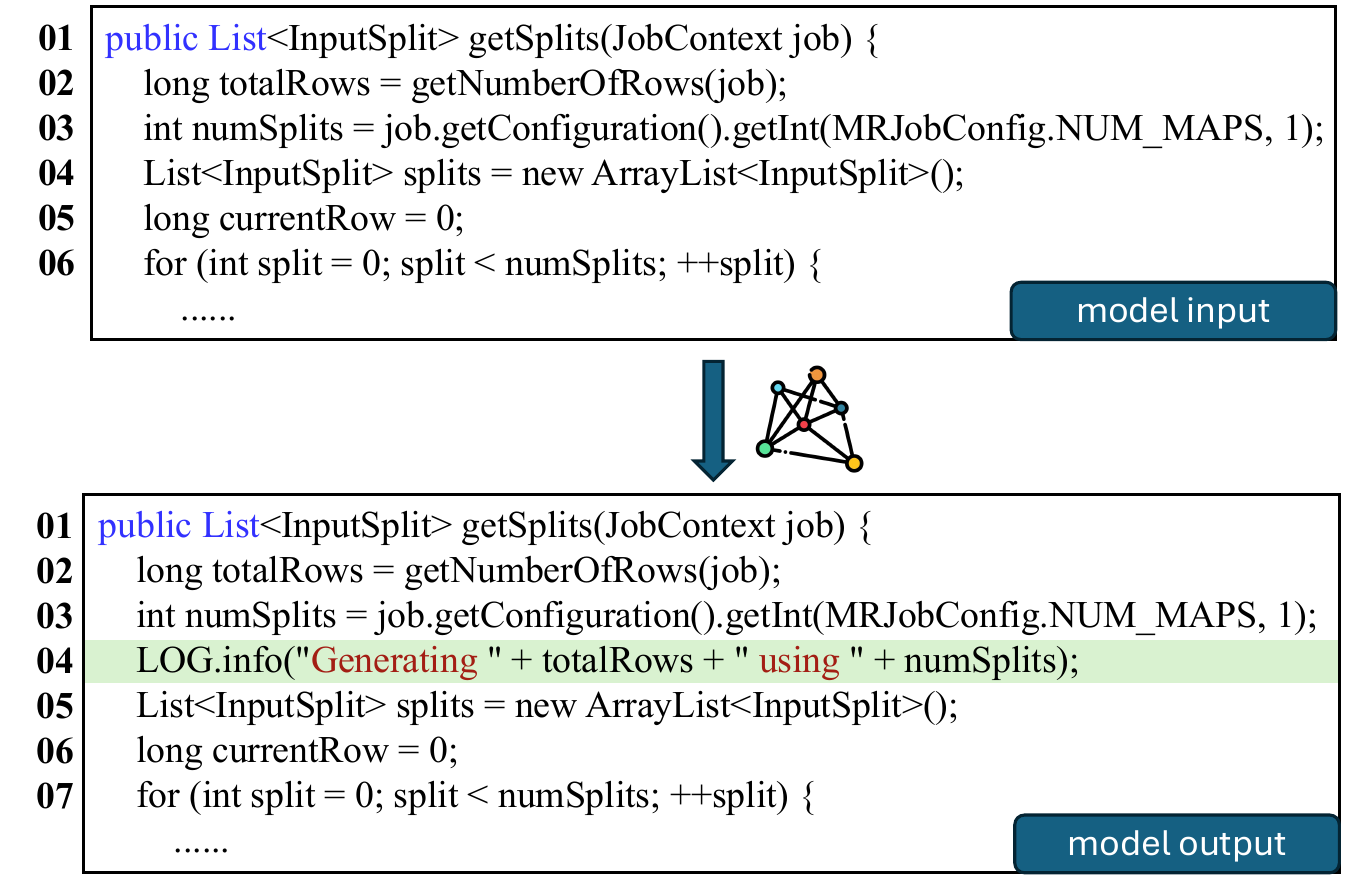}
    \caption{Task formulation: given a method which missing a logging statement, the model is asked to automated generate a logging statement.}
    \label{fig:task}
\end{figure}

This paper focuses on the automated logging task (i.e., where-to-log + what-to-log), which to some extent can be viewed as a code editing problem: when presented with lines of code, usually corresponding to a method, the generator's task is to identify both the precise location for logging, referred to as the logging point, and the complete logging statement (i.e., level, variables, and text). The predicted logging point should match the one that was originally present in the source file before being removed, and the predicted logging statement itself should closely resemble the excised original. Figure~\ref{fig:task} provides a visual example of this task, showing how a proficient logging statement generator would intelligently incorporate \texttt{LOG.info("Generating " + totalRows + " using " + numSplits);} at line 4. It is important to highlight that this task is distinctly separate from the comprehensive empirical investigation conducted by Li et al.~\cite{li2023exploring}, which predominantly examines the question of what-to-log but lacks the consideration of where-to-log.

\subsection{Large Language Models}
The evolution of language models in recent times can be divided into three transformative phases. Initially, there were neural language models (NLMs), followed by the phase of pre-trained language models (PLMs), and the current era sees the prominence of LLMs. Pre-trained models such as CodeT5~\cite{wang2021CodeT5} and PLBART~\cite{ahmad2021unified} have achieved noteworthy success in software engineering applications primarily due to task-specific pre-training processes. However, LLMs have brought about a revolutionary change in the field due to their immense parameter counts, often exceeding 10 billion, and their comprehensive pre-training data. \added{These models, unlike their pre-training predecessors, exhibit emergent capabilities that allow them to achieve robust performance across a wide range of tasks without necessitating fine-tuning tailored to specific tasks~\cite{geng2024large}.} This quality substantially diminishes the requirement for resource-heavy training sessions. Within the realm of software engineering, LLMs are mainly categorized into two groups. Unified large language models, such as GPT-4o and LLaMA, which are designed to integrate natural language and code corpora, whereas code-specific large language models, like StarCoder~\cite{li2023starcoder} and CodeLlama~\cite{roziere2023code}, are developed for specialization in tasks centered around coding.

\added{Methodologies for leveraging these models typically follow two distinct paradigms based on model scale and accessibility. The first is the prompt-based paradigm, which exploits the zero-shot or few-shot capabilities of massive-scale LLMs (e.g., GPT-4). This approach relies on carefully engineered prompts, using techniques like in-context learning (ICL) and chain-of-thought (CoT) to guide the model without updating its parameters. This paradigm is characteristic of the largest models, which often pose significant resource challenges and are typically accessed via APIs.}

\added{A contrasting paradigm involves the tuning-based adaptation of smaller-scale, open-source models through PEFT techniques like Low-Rank Adaptation (LoRA). This approach has given rise to a focus on what this paper terms SOLMs, which are designed to balance performance with practical deployability. While the exact parameter threshold for such models is an evolving topic, they are often distinguished from massive-scale LLMs by a specific size cap, such as under 10 billion parameters~\cite{yang2024chain,fu2023specializing}. In this study, we adopt a hardware-centric threshold and define SOLMs as models with fewer than 14 billion parameters. This practical boundary reflects the limit of what can be efficiently fine-tuned and deployed on a single, widely available high-end GPU (e.g., an NVIDIA A100 80G), a rationale similarly employed for defining lightweight models in other software engineering contexts~\cite{yang2024chain}. This distinction is critical, as SOLMs represent a class of models that can be leveraged by enterprises to create specialized, cost-effective, and privacy-preserving solutions for tasks like automated logging, circumventing the resource demands of their larger counterparts.}

\subsection{LLM Applications in Software Engineering Task}

Researchers have conducted in-depth investigations into the use of Large Language Models (LLMs) in a variety of software engineering tasks, including but not limited to code completion~\cite{izadi2022codefill,izadi2024language}, vulnerability detection~\cite{sun2024llm4vuln,zhao2024coding}, program repair~\cite{wei2023copiloting,lin2024one}, and test generation~\cite{chen2024chatunitest,liu2024testing}. These studies highlight the versatility and adaptability of LLMs in effectively tackling a range of software engineering challenges.

In certain tasks, approaches that utilize prompts have been shown to yield superior outcomes~\cite{imani2024context,jiang2024lilac}. Pan et al., for instance, studied the efficiency of LLMs in the context of code translation~\cite{pan2024lost}. Within the range of models assessed, which encompassed both SOLMs and GPT-4, the top-performing SOLM, known as StarCoder, managed to reach a successful translation rate of 14.5\%, in contrast to the success rate of 47.3\% achieved by GPT-4. In contrast, SOLMs have been shown to achieve comparable outcomes in specific domains. For instance, Tian et al.~\cite{tian2024Large} observed an F1-score of 86.58\% using UniXCoder for the task of detecting equivalent mutants, providing a notable contrast to the performance of GPT-4, which recorded an F1-score of approximately 55.90\%. Furthermore, when employing Vicuna 13B in conjunction with the innovative LogPrompt strategy, the performance was found to be on par with that of GPT-4, as reported in~\cite{liu2024interpretable}.

\section{Experimental Design}

\begin{figure}
    \centering
    \includegraphics[width=0.8\linewidth]{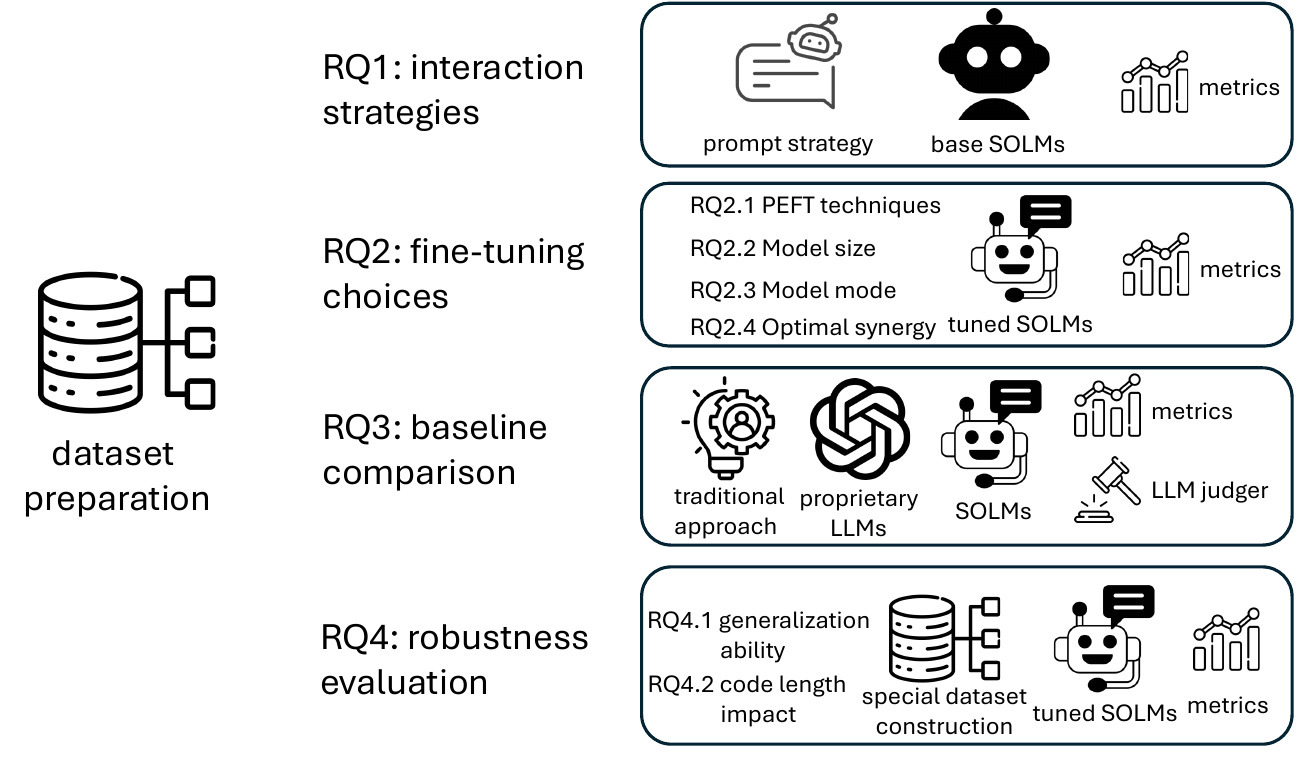}
    \caption{The overview of our experimental design with four research questions.}
    \label{fig:framework}
\end{figure}

Figure~\ref{fig:framework} illustrates the overview of our study design. Initially, based on the AL-Bench dataset~\cite{tan2025ALBench}, we construct the fine-tuning, valid and test datasets. Then we investigate SOLMs for automated logging through four research questions (RQs). RQ1 investegates the most effective prompt strategies for using SOLMs in this task. RQ2 aims to determine the optimal fine-tuning strategies by PEFT techniques, model sizes, and model types. Following that, RQ3 compares the performance of fine-tuned SOLMs against existing methods and LLM baselines. Finally, RQ4 assesses the ability of SOLMs to generalize their logging statement generation capabilities across diverse unseen code repositories.

\subsection{Dataset Preparation}

\begin{table}[tbp]
\centering
\caption{Statistics of source repositories of AL-Bench~\cite{tan2025ALBench}.}
\label{tab:dataset_statistics}
\begin{tabular}{@{}lllcccc@{}}
\toprule
\textbf{Index} & \textbf{Dataset}     & \textbf{Domains}     & \textbf{\#Stars}                & \textbf{\#Forks}                                & \textbf{\#LOLS}                 \\ \midrule
R1 & Dbeaver        &  Database Management  & 43.3k                & 3.7k                              & 1.7k                 \\
R2 & Dolphinscheduler  & Task Scheduling & 13.5k                & 4.8k                              & 1.9k                     \\
R3 & Doris           & High Performance Database & 13.6k                & 3.4k                                         & 2.9k                     \\
R4 & Flink          & Data Processing  & 24.8k                     & 13.5k                                       & 3.2k                     \\
R5 & Hadoop        & Distributed Storage   & 15.1k                     & 9.0k                                          &  16.0k                    \\
R6 & Kafka        & Messaging Systems    & 30.0k                     & 14.3k                                         &       3.4k               \\
R7 & Keycloak    & Identity and Access Management     & 26.9k                     & 7.2k                                        &     1.0k                 \\
R8 & Pulsar      &  Messaging Systems     & 14.6k                     & 3.6k                                        &     7.2k                 \\
R9 & Thingsboard   &  IoT Platform   & 18.7k                     & 5.5k                                      &    2.7k                  \\
R10 & Zookeeper     & Distributed Coordination   & 12.5k                     & 7.3k                     &                        1.9k                 \\ \bottomrule
\end{tabular}
\end{table}

\subsubsection{Studied dataset.}
\added{To evaluate the performance of automated logging, we selected AL-Bench~\cite{tan2025ALBench}, the most recently proposed large-scale benchmark for this task. Our choice was deliberate, as AL-Bench offers several critical advantages over prior datasets that are essential for a rigorous evaluation of SOLMs and LLMs.}

\added{First, AL-Bench is specifically designed to evaluate the end-to-end task of logging statement generation, encompassing both the placement (where-to-log) and content creation (what-to-log) aspects. Second, the benchmark was constructed using stringent quality criteria (e.g., >=10k stars, >=1k logging statements, >=500 log-related issues per project), sourcing data exclusively from 10 popular, well-maintained Java projects across a diverse set of domains, including database management, task scheduling, and IoT platforms. This ensures that the ground truth logging statements are of high quality and that our evaluation is representative of real-world practices.}

\added{Most importantly, AL-Bench was developed with a keen awareness of the challenges posed by LLMs, particularly the threat of data leakage. The benchmark's creators employed specific mitigation strategies to minimize the likelihood that its content was included in the pre-training corpora. For instance, all code snippets were processed with the Google-Java-Format tool and wrapped in a generic class, a method designed to alter the code's original structure and appearance. This thoughtful step, combined with the use of the most recent project versions, further reduces the risk that a model's performance might be inflated by prior exposure to the test data.}

\added{The thoughtful design of AL-Bench provides a robust foundation for evaluating modern language models. To ensure a consistent and fair comparison across all methods, we adopted AL-Bench as the sole benchmark for our experiments. This unified approach allows us to rigorously evaluate our fine-tuned models alongside existing state-of-the-art tools within the same controlled environment. Table~\ref{tab:dataset_statistics} provides further statistical details on these source repositories.}

\subsubsection{Pre-processing and dataset construction.}
\added{To construct the primary dataset used for our main experiments (RQ1-RQ3, RQ4.1), we performed several pre-processing steps on the original AL-Bench dataset. Firstly, to ensure a fair comparison and accommodate the input constraints of certain baseline models, we established a maximum input length of 512 tokens. Data instances with code snippets exceeding this threshold were set aside for a separate analysis.} Secondly, the original construction of AL-Bench could generate multiple data points from a single Java function if it contained multiple logging statements, with each data point representing one specific automated logging case. To prevent potential data leakage, where highly similar code snippets from the same source file might inadvertently appear across different dataset splits (e.g., fine-tuning and testing), we implemented a file-level splitting strategy. This approach ensures that all data instances originating from the same Java file are strictly allocated to only one of the fine-tuning, validation, or test sets. After applying these pre-processing steps, we obtained a final dataset comprising 33,224 instances. We then partitioned this dataset into fine-tuning, validation, and test sets, targeting an 8:1:1 ratio. Because our file-level splitting strategy required keeping all instances from a single file within the same set, the resulting distribution was approximate. The final fine-tuning set contains 26,713 instances, the validation set contains 3,508 instances, and the test set contains 3,003 instances.

\added{In addition, to specifically investigate the impact of code length in RQ4.2, we utilized the previously set-aside longer code snippets to construct a specialized test set. This test set is also going through the above pre-processing steps. The detailed motivation and use of this set are further elaborated in the approach part of RQ4.2.}


\subsection{Studied Models}
In our study, we investigate the performance of the following SOLMs for logging statement generation. These models have been widely adopted in the literature related to SE tasks, including:

\begin{itemize}[leftmargin=*]
    \item \textbf{LLaMA3}~\cite{grattafiori2024llama} is Meta's latest LLM and refines the LLaMA 2 framework. It stands as a prominent open source LLM used in numerous software applications. Trained on an extensive and varied dataset far surpassing its predecessor, LLaMA 3 exhibits significantly improved proficiency in reasoning, code generation, and instruction adherence.
    \item \textbf{Mistral}~\cite{jiang2023mistral7b} is noted for its efficiency and performance, utilizing Grouped-Query Attention (GQA) and Sliding Window Attention (SWA) for faster inference and broader context handling, and demonstrates robust general abilities and notable coding skills.
    \item \textbf{CodeLlama}~\cite{roziere2023code} is a series of LLMs specialized in generating and completing code, based on the LLaMA2 framework. These models are initially trained using a dataset of 500 billion code tokens and subsequently refined to manage extended context effectively.
    \item \textbf{Qwen2.5-coder}~\cite{hui2024qwen2} is a code-specialized version of the Qwen2.5~\cite{qwen2} family, which inherits Qwen's multilingual capabilities and architectural improvement. While demonstrating strong and comprehensive coding abilities, it also possesses good general and mathematical skills.
\end{itemize}

\subsection{Baselines}
\label{baselines}
\added{To evaluate SOLMs performance, we select the baselines by conducting a literature review of relevant papers published in SE venues. From this, we find the following methods for evaluation: \textbf{LANCE}~\cite{mastropaoloUsingDeepLearning2022}, \textbf{LEONID}~\cite{mastropaolo2024log}, \textbf{UniLog}~\cite{xu2024unilog}, \textbf{FastLog}~\cite{xie2024fastlog}, and \textbf{SCLogger}~\cite{li2024go}.}
Additionally, we examine several LLM baselines, including general-purpose LLMs (\textbf{Claude3.7-sonnet}, \textbf{GPT4o}, \textbf{LLaMA3.1-405b}), and a code-specific LLM (\textbf{Deepseek-coder-v3}). 

In relation to the methodologies for prompting, we derive our approach from the study~\cite{gao2024search} and incorporate four different strategies: instruction prompting (\textbf{base}), in-context learning (\textbf{ICL}),  retrieval-augmented generation (\textbf{RAG}) and chain of thought (\textbf{CoT}). The instruction prompting strategy involves directly prompting LLMs to generate logging statements using identical inputs as those provided to SOLMs, without any supplementary data. The ICL approach consists of providing one random example before the main query to assist the model in grasping the nature of the task more effectively. 
\added{In the RAG approach, we enhance the prompt by providing the most similar example to guide the model. Instead of selecting an example randomly as in ICL, we retrieve the most similar instance from our validation set to serve as a one-shot demonstration. This process leverages the validation set as a dedicated knowledge pool. Specifically, for each input sample from the test set, we employ the BM25 algorithm~\cite{robertson2009probabilistic} to identify and retrieve the single most similar code snippet from the validation set. This retrieved example is then prepended to the main query within the prompt.}
\added{Finally, for the CoT strategy, we employ a zero-shot approach that explicitly instructs the model to "reason step-by-step" about the purpose, placement, content, and level of the logging statement before providing the final code output.} The details of the prompts are shown in Figure~\ref{fig:prompt}.

\subsection{Strategies for Parameter-Efficient Fine-Tuning}
We examine how the following PEFT strategies influence the performance of SOLMs when automated logging.

\underline{Prefix tuning}~\cite{li2021prefix}, is a PEFT strategy designed to adapt LLMs to specific downstream tasks while keeping the original model parameters entirely frozen. Instead of modifying the model's weights, it introduces a small set of trainable continuous vectors, known as the "prefix", which are prepended to the key and value sequences within the multi-head attention mechanisms of the Transformer architecture, typically applied to the topmost $L$ layers. Specifically, for a given layer $l$, trainable prefix matrices $\mathbf{P}_k \in \mathbb{R}^{K \times C}$ and $\mathbf{P}_v \in \mathbb{R}^{K \times C}$ (where $K$ is the prefix length, a key hyperparameter, and $C$ is the hidden dimension) are concatenated with the original key ($\mathbf{K}_l \in \mathbb{R}^{M \times C}$) and value ($\mathbf{V}_l \in \mathbb{R}^{M \times C}$) matrices derived from the $M$ input tokens, forming augmented matrices $\mathbf{K}'_l = [\mathbf{P}_k; \mathbf{K}_l]$ and $\mathbf{V}'_l = [\mathbf{P}_v; \mathbf{V}_l]$. During fine-tuning, only the parameters comprising these prefix matrices ($\mathbf{P}_k, \mathbf{P}_v$ across the selected layers) are optimized via gradient descent, learning task-specific representations that effectively steer the frozen model's attention and subsequent computations towards generating appropriate outputs for the target task. This approach significantly reduces the number of trainable parameters compared to full fine-tuning, requires storing only the small prefix parameters per task, and avoids catastrophic forgetting by leaving the core model untouched.

\underline{Prompt tuning}~\cite{lester2021power} offers an even more lightweight approach by confining trainable parameters exclusively to continuous prompt embeddings added only at the input layer, while freezing the entire pre-trained model, including its word embedding table. This method prepends a sequence of $K$ learnable prompt embeddings, represented by a single trainable matrix $\mathbf{P}_{\text{emb}} \in \mathbb{R}^{K \times C}$ (where $K$ is the prompt length and $C$ is the model's embedding dimension), to the original sequence of $M$ input token embeddings $\mathbf{E} \in \mathbb{R}^{M \times C}$, yielding an augmented input sequence $\mathbf{E}' = [\mathbf{P}_{\text{emb}}; \mathbf{E}]$. This combined sequence $\mathbf{E}'$ is then fed directly into the first layer of the frozen Transformer backbone. During the fine-tuning process, only the parameters of the prompt embedding matrix $\mathbf{P}_{\text{emb}}$ are updated. The core idea is that these learned continuous vectors act as a "soft prompt" or task instruction, conditioning the frozen model's behavior without requiring any internal modifications. Prompt Tuning demonstrates significant efficiency regarding parameter usage, typically necessitating the update of fewer than 0.1\% of the total model parameters. This characteristic renders it highly efficient in terms of both storage and computation, especially in multi-task contexts. 

\added{\underline{LoRA}}~\cite{hu2022lora} provides a distinct PEFT mechanism based on the hypothesis that the change in weights during model adaptation has a low intrinsic rank. Instead of adding prefix or prompt tokens, LoRA freezes the original pre-trained weights $\mathbf{W}_0 \in \mathbb{R}^{d \times k}$ of selected layers (commonly the query, key, value, and output projection matrices in self-attention, and sometimes feed-forward layers) and injects trainable, rank-decomposition matrices in parallel. Specifically, the weight update $\Delta \mathbf{W}$ is approximated by the product of two smaller, low-rank matrices: $\mathbf{W}_{\text{down}} \in \mathbb{R}^{d \times r}$ and $\mathbf{W}_{\text{up}} \in \mathbb{R}^{r \times k}$, where the rank $r$ is a crucial hyperparameter significantly smaller than the original dimensions ($r \ll \min(d, k)$). \added{The modified forward pass for an input $x$ computes the output by merging the original and adapter paths: $h_{\text{adapted}} = \mathbf{W}_0 x + \Delta \mathbf{W} x = \mathbf{W}_0 x + \alpha (\mathbf{W}_{\text{down}} \mathbf{W}_{\text{up}}) x$. In this formulation, $\alpha$ is a scaling hyperparameter. In some implementations, this scaling factor is set in proportion to the rank $r$ (e.g., as $\frac{\alpha}{r}$), which helps stabilize the adaptation process by ensuring the magnitude of the update remains consistent when experimenting with different values of $r$. }During fine-tuning, only the parameters of $\mathbf{W}_{\text{down}}$ and $\mathbf{W}_{\text{up}}$  are optimized. A significant advantage of LoRA is that the learned adapter weights can be mathematically merged with the original weights after training, resulting in a single weight matrix per adapted layer and incurring zero additional inference latency compared to the original model, while still offering substantial parameter savings during training and allowing easy task switching by loading different adapter pairs.

\added{\underline{QLoRA}}~\cite{dettmers2023qlora} represents a significant advancement in memory-efficient fine-tuning, specifically designed to make the adaptation of extremely large language models feasible on commodity hardware with limited VRAM. It ingeniously combines low-precision quantization of the base model with the LoRA technique. The core strategy involves loading the massive pre-trained base model $\mathbf{W}_0$ with its weights quantized to a very low bit-format, most notably 4-bit NormalFloat (NF4), a data type empirically shown to be effective for normally distributed weights, and keeping these quantized weights $Q(\mathbf{W}_0)$ frozen. Standard LoRA adapters, consisting of low-rank matrices $\mathbf{W}_{\text{down}}$ and $\mathbf{W}_{\text{up}}$, are then introduced parallel to these quantized layers, but crucially, these adapter weights are maintained and trained in a higher precision format, typically BFloat16, to preserve adaptation capacity. The forward pass thus involves computations using the low-precision base model and the higher-precision adapters: $\mathbf{h}_{\text{adapted}} \approx Q(\mathbf{W}_0) \mathbf{x} + \alpha (\mathbf{W}_{\text{down}} \mathbf{W}_{\text{up}}) \mathbf{x}$. To further minimize memory bottlenecks, QLoRA incorporates innovations like double quantization and paged optimizers. By drastically reducing the memory footprint of the base model weights, activations (due to lower precision), and optimizer states, QLoRA enables fine-tuning models with tens or hundreds of billions of parameters on single consumer GPUs, while aiming to retain the task performance levels achieved by full-precision LoRA.

\subsection{Evaluation method}

\subsubsection{Tradictional evaluating metrics}

Considering earlier research~\cite{li2023exploring,tan2025ALBench}, we assess the performance of automated generation of logging statements by focusing on four aspects: the logging point, the logging levels, the logging text, and the logging variables. While each of these components highlights distinct aspects of system runtime information, they collectively serve as essential and complementary resources that aid engineers in analyzing and understanding system behaviour.

\underline{Logging point:} We use position accuracy (PA) to evaluate the performance of logging location. To quantify PA, we compare the predicted locations of logging statements against their ground truth positions in the source code. This metric is formally defined as the ratio of correctly positioned logging statements ($LS_{position\_correct}$) to the total number of logging statements ($LS_{all}$), expressed as $\frac{LS_{position\_correct}}{LS_{all}}$.

\underline{Logging level:}
We use the Level Accuracy (LA) and Average Ordinal Distance Score (AOD) for evaluating the prediction of logging levels. Given the significant semantic differences between these levels and their implications for system monitoring and maintenance, we rigorously assess LA by comparing predicted log levels against their ground truth values in the source code. This metric is formally defined as the ratio of correctly predicted log levels ($LS_{level\_correct}$) to the total number of logging statements ($LS_{all}$), expressed as $\frac{LS_{level\_correct}}{LS_{all}}$.
AOD evaluates how closely the current logging level aligns with the recommended logging level for each specific logging statement, as detailed in~\cite{liDeepLVSuggestingLog2021}. The formula to calculate AOD is given by: $AOD = \frac{\sum_{i=1}^{N} (1 - \frac{Dis(a_i, s_i)}{MaxDis(a_i)})}{N}$, where $N$ represents the total number of logging statements in consideration. The term $MaxDis(a_i)$ is used to denote the maximum potential distance for the actual log level $a_i$. \added{We verify that all 10 projects within our dataset consistently utilize the same five logging levels (trace, debug, info, warn, and error), making a unified distance appropriate for this AOD metric.}

\underline{Static logging text:} Same as previous study~\cite{li2023exploring, mastropaoloUsingDeepLearning2022,dingLoGenTextAutomaticallyGenerating2022}
, our evaluation of static logging texts is conducted through the application of two metrics commonly employed in the domain of machine translation: BLEU~\cite{papineni2002bleu} and ROUGE~\cite{lin2004rouge}. These metrics, grounded in n-gram analysis, are instrumental in assessing the degree of similarity between log messages that are generated computationally and those authored by developers. They provide a normalized score continuum from 0 to 1, with elevated scores indicative of a closer resemblance. In our methodology, we specifically implement various forms of these metrics, identified as BLEU-4 and ROUGE-L.

\underline{Dynamic logging variables:}
We use Precisely Match Rate (PMR) and F1 to evaluate dynamic logging variables. 
PMR ensures consistency in the capture of variable runtime data—a critical aspect of log effectiveness. We extract dynamic variables from both reference implementations and predicted logging statements, then perform exact matching to evaluate correspondence. PMR is formally defined as the ratio of exactly matched dynamic variables ($LS_{variable\_correct}$) to the total number of logging statements ($LS_{all}$), expressed as $\frac{LS_{variable\_correct}}{LS_{all}}$.
Moreover, consider each logging statement and let us define $S_{ud}$ as the set of variables included in the generated logging statement, while $S_{gt}$ pertains to the set of variables present in the actual logging statement. In our analysis, we determine and present the following metrics: the precision, which is the ratio of variables from the updates that accurately match those in the actual logging ($precision = \frac{S_{ud} \cap S_{gt}}{S_{ud}}$); the recall, which indicates the fraction of actual variables correctly anticipated by the model ($recall = \frac{S_{ud} \cap S_{gt}}{S_{gt}}$); and, lastly, their harmonic mean, expressed as the F1 score ($F_1 = 2*\frac{Precision*Recall}{Precision+Recall}$)~\cite{li2023exploring}.

\subsubsection{LLM-as-a-judge}
The evaluation of automatically generated logging statements, drawing upon our experimental findings and established prior work, conventionally proceeds by assessing distinct modules such as placement, verbosity level, and textual content. However, it is increasingly evident that certain prevalent metrics do not adequately capture the nuanced performance aspects of generated logging statements~\cite{roy2021reassessing,gros2020code}. For instance, in the assessment of static text components, metrics like BLEU-4 and ROUGE-L are confined to lexical similarity, largely overlooking crucial semantic congruity. This limitation presents a formidable challenge in establishing a unified and robust methodology for evaluating the overall quality of automatically generated logging statements.

Recently, research within the NLP domain has explored the application of LLM to appraise the quality of LLM-generated content, known as "LLM-as-a-judge". While human evaluation remains a reliable way, its inherent drawbacks of being time-consuming and cost-intensive run counter to the objectives of automated evaluation. Consequently, researchers are increasingly investigating methods to prompt or train LLMs to align with human evaluative preferences, thereby offering a scalable alternative to manual assessment. Supporting this direction, an empirical study by Wang et al.~\cite{wang2025can} has demonstrated the efficacy of the LLM-as-a-judge approach across various SE tasks. Their findings indicate that output-based evaluation methods, when coupled with state-of-the-art LLMs, yield optimal performance irrespective of the specific inference strategies employed. Informed by these advancements, this paper adopts the LLM-as-a-judge methodology to augment the quality assessment of automatically generated logging statements.

Specifically, we select three LLMs recognized for their superior performance in code-related tasks: Claude3.7-Sonnet, Deepseek-coder-v3, and GPT-4o. We choose this model due to their robust code comprehension and generation capabilities, which are critical for assessing the nuanced quality of logging statements. The evaluation process involves providing each LLM judge with the input code context, the generated logging statement from model, and the corresponding ground truth logging statement. The judges assign scores ranging from 0 to 3, where higher scores indicate greater alignment with the ground truth in terms of logging point accuracy, level appropriateness, static text quality, and dynamic variable correctness. For each generated logging statement, the final score is the average of the scores provided by these three LLM judges. To ensure consistency and reliability, we develop a comprehensive scoring guideline, which outline specific criteria for evaluating each component of the logging statement. These criteria address syntactic accuracy, semantic relevance, and contextual appropriateness, mitigating the limitations of traditional metrics like BLEU-4 and ROUGE-L.

\begin{tcolorbox}[boxsep=1pt,left=2pt,right=2pt,top=3pt,bottom=2pt,width=0.95\columnwidth,colback=white,boxrule=1pt, colframe=black, colbacktitle=white!30!black,toptitle=2pt,bottomtitle=1pt,opacitybacktitle=0.4, sharp corners,center]

\textbf{Score Guideline}

\textbf{0: (Unacceptable)} The logging statement is syntactically incorrect or misplaced. Formatting deviates significantly, and code changes may impair functionality or maintainability.

\textbf{1: (Significant Issues)} The statement is syntactically correct and appropriately placed but has major flaws: vague static text, incorrect log level, or missing key variables. Formatting inconsistencies or minor code alterations reduce readability but preserve functionality.

\textbf{2: (Mostly Correct with Minor Flaws)} The statement is semantically accurate, includes most relevant details, and uses an appropriate log level. Minor issues include verbose text, suboptimal formatting, or slight stylistic deviations. Functionality is preserved with trivial changes.

\textbf{3: (Highly Accurate)} The statement is precise, concise, and matches the ground truth. It uses the correct log level, parameterized logging, and consistent styling. The code retains full functionality and maintainability.

\end{tcolorbox}

\subsection{Implimentation Details}
To evaluate conventional logging approaches and LLMs, we reproduced conventional methods using replication packages provided by their authors. 
\added{For LLMs, we generated logging statements by calling their official APIs. To promote deterministic outputs, we set the temperature to 0. We acknowledge that for some proprietary services like OpenAI, setting the temperature to 0 alone does not guarantee reproducibility~\cite{OpenAI2025Reproducible}.} 

\added{For the fine-tuning of SOLMs, we employed the Axolotl framework. The training objective was a standard causal language modeling task, where the model learns to predict the next token in a sequence. We utilized Cross-Entropy Loss as the objective function to minimize during training.
The training data was carefully prepared to align with our prompting strategies. For each data instance, the input source code was formatted using a specific prompt template (e.g., Base, RAG, or CoT) to create the final input sequence. The ground-truth code containing the logging statement served as the target label.}
\added{All SOLMs were fine-tuned for one epoch. We use a learning rate of 1e-4 with a cosine scheduler, a global batch size of 64, and the Adam optimizer. For LoRA and QLoRA, we used a rank (r) of 16 and an alpha of 32. All experiments, including training, fine-tuning, and inference, were conducted on a single NVIDIA A100 80GB GPU provided by Modal~\cite{modal}, a serverless cloud infrastructure platform. To ensure full reproducibility, the complete scripts, configuration files, and data preparation details are publicly available in our replication package~\cite{artifacts}.}

\section{Empircial Study Results}
\label{results}

\subsection{RQ1: What are the most effective prompting strategies for using SOLMs in automated logging?}

\textbf{Motivation.}
Initially, our objective is to determine if the manner in which we engage with the model during the inference phase can have a profound effect on its success in generating automated logging statements. The structuring of the input prompt plays a crucial role in shaping how the SOLM interprets the given task and how it subsequently produces its outputs.

\begin{figure}
    \centering
    \includegraphics[width=0.75\linewidth]{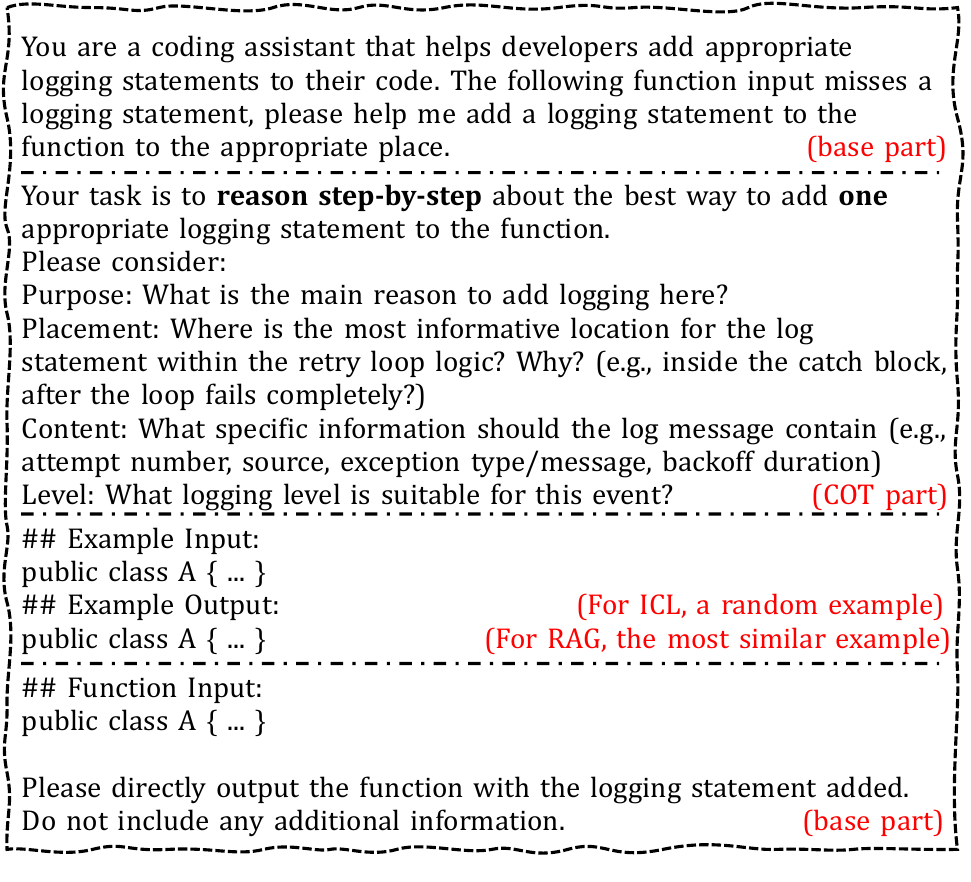}
    \caption{The prompt template for automated logging.}
    \label{fig:prompt}
\end{figure}

\textbf{Approach.}
In addressing this question, our study is focused on evaluating the effectiveness of current prompting techniques within the domain of log generation. We specifically analyze, based on the definitions laid out in Section~\ref{baselines}, the effectiveness of several prompting strategies: basic instruction prompting (\textbf{base}), in-context learning (\textbf{ICL}), retrieval-augmented generation (\textbf{RAG}), and chain-of-thought (\textbf{CoT}). The specific details regarding the prompt templates used can be found in Figure~\ref{fig:prompt}. To strengthen the generalizability of our findings, we employ multiple 7B instruction-following models, namely LLaMA3, Mistral, CodeLlama, and Qwen2.5-Coder, all in their original configurations.

\textbf{Results.}
The quantitative evaluation comparing the four prompting techniques across the four selected 7B SOLMs is presented in Table~\ref{tab:rq1}. Our analysis of these results yields two primary findings concerning the performance of un-fine-tuned models and the efficacy of different prompt strategies for automated logging generation.

First, we observe a distinct performance disparity between the general-purpose instruction-following models (LLaMA3, Mistral) and the code-specific models (CodeLlama, Qwen2.5-coder) when used without any task-specific fine-tuning, i.e., the general-purpose models generally demonstrate superior performance on logging. For instance, LLaMA3 and Mistral achieve peak PA scores of 21.01 and 18.15, respectively (both using RAG), substantially higher than the peak PA scores achieved by CodeLlama (11.62 with ICL) and Qwen2.5-coder (13.25 with CoT). We attribute this trend to the weaker instruction-following capabilities in the original code-specific models. Furthermore, upon analyzing the failure cases, we noted a higher tendency for CodeLlama and Qwen2.5-coder to refuse the prompt or return empty responses compared to LLaMA3 and Mistral, which negatively impacts their effectiveness in this experimental setting. This suggests that, without fine-tuning, the broader instruction comprehension of general-purpose models may be more advantageous for automated logging tasks.

\begin{tcolorbox}[boxsep=1pt,left=2pt,right=2pt,top=3pt,bottom=2pt,width=\columnwidth,colback=white!95!black,boxrule=0pt, colbacktitle=white!30!black,toptitle=2pt,bottomtitle=1pt,opacitybacktitle=0.4, sharp corners]

\textbf{Findings 1}: Without fine-tuning, general-purpose models outperform code-specific models for automated logging due to their better instruction-following ability.

\end{tcolorbox}

Second, RAG emerges as the most effective prompting technique for enhancing automated logging generation performance. While other techniques occasionally yield the top score for an isolated metric, RAG demonstrates the most significant and robust improvements across the majority of metrics and models. Notably, for Mistral, RAG achieves the highest scores across all evaluated metrics, including PA (18.15), F1 (40.21), and ROUGE-L (34.74). For LLaMA3, RAG secures the top performance in PA (21.01), PMR (41.68), F1 (46.28), BLEU-4 (15.95), and ROUGE-L (36.73). For CodeLlama and Qwen2.5-coder, RAG generally leads to substantial gains over the baseline, particularly in metrics like F1 (CodeLlama: 39.01, Qwen2.5-coder: 49.28) and ROUGE-L (CodeLlama: 34.36, Qwen2.5-coder: 37.57). These results strongly indicate that providing relevant contextual information retrieved from a knowledge base significantly aids the SOLMs in accurately determining where to log, what variables to include, and formulating appropriate log messages, making RAG a highly promising strategy.

\begin{tcolorbox}[boxsep=1pt,left=2pt,right=2pt,top=3pt,bottom=2pt,width=\columnwidth,colback=white!95!black,boxrule=0pt, colbacktitle=white!30!black,toptitle=2pt,bottomtitle=1pt,opacitybacktitle=0.4, sharp corners]

\textbf{Findings 2}: RAG proves the most effective prompting technique, significantly enhancing automoated logging statement generation performance across models and metrics.

\end{tcolorbox}


\begin{table}[t]
\centering
\caption{Comparison of prompting techniques for automated logging generation using four 7B instruction-following SOLMs. The greener, the better.}
\label{tab:rq1}
\setlength{\tabcolsep}{4pt}
\begin{tabular}{@{}ll||c||cc|cc|cc@{}}
\toprule
\multirow{2}{*}{Model} & \multirow{2}{*}{Techs} & Location & \multicolumn{2}{c}{Level} & \multicolumn{2}{c}{Variable} & \multicolumn{2}{c}{Text} \\ \cmidrule(l){3-9}
& & PA & LA & AOD & PMR & F1 & BLEU-4 & ROUGE-L \\ \midrule
\multirow{4}{*}{LLaMA3} 
& base & \heatcell{48}{13.32} & \heatcell{100}{54.50} & \heatcell{100}{83.06} & \heatcell{88}{38.00} & \heatcell{41}{30.41} & \heatcell{30}{9.68} & \heatcell{41}{28.92} \\
& ICL  & \heatcell{26}{9.99} & \heatcell{85}{50.67} & \heatcell{83}{81.25} & \heatcell{63}{30.00} & \heatcell{43}{30.95} & \heatcell{48}{11.31} & \heatcell{54}{30.60} \\
& RAG  & \heatcell{100}{21.01} & \heatcell{97}{53.72} & \heatcell{91}{82.09} & \heatcell{100}{41.68} & \heatcell{100}{46.28} & \heatcell{100}{15.95} & \heatcell{100}{36.73} \\
& COT  & \heatcell{0}{6.16} & \heatcell{0}{28.65} & \heatcell{0}{72.64} & \heatcell{0}{10.27} & \heatcell{0}{19.15} & \heatcell{0}{6.96} & \heatcell{0}{23.49} \\ \midrule
\multirow{4}{*}{Mistral} 
& base & \heatcell{54}{12.85} & \heatcell{26}{46.89} & \heatcell{30}{79.86} & \heatcell{49}{25.13} & \heatcell{14}{23.70} & \heatcell{30}{9.91} & \heatcell{11}{27.19} \\
& ICL  & \heatcell{47}{11.99} & \heatcell{0}{40.56} & \heatcell{0}{77.08} & \heatcell{74}{31.11} & \heatcell{40}{28.78} & \heatcell{0}{8.05} & \heatcell{0}{26.26} \\
& RAG  & \heatcell{100}{18.15} & \heatcell{100}{65.14} & \heatcell{100}{86.35} & \heatcell{100}{37.25} & \heatcell{100}{40.21} & \heatcell{100}{14.36} & \heatcell{100}{34.74} \\
& COT  & \heatcell{0}{6.63} & \heatcell{39}{50.25} & \heatcell{23}{79.23} & \heatcell{0}{13.57} & \heatcell{0}{21.02} & \heatcell{47}{11.01} & \heatcell{48}{30.35} \\ \midrule
\multirow{4}{*}{CodeLLAMA} 
& base & \heatcell{29}{6.99} & \heatcell{66}{57.14} & \heatcell{67}{84.29} & \heatcell{100}{37.62} & \heatcell{66}{35.74} & \heatcell{63}{12.21} & \heatcell{61}{31.71} \\
& ICL  & \heatcell{100}{11.62} & \heatcell{5}{48.14} & \heatcell{0}{79.87} & \heatcell{35}{26.36} & \heatcell{0}{29.47} & \heatcell{0}{9.50} & \heatcell{0}{27.54} \\
& RAG  & \heatcell{47}{8.16} & \heatcell{100}{62.04} & \heatcell{100}{86.43} & \heatcell{62}{31.02} & \heatcell{100}{39.01} & \heatcell{100}{13.82} & \heatcell{100}{34.36} \\
& COT  & \heatcell{0}{5.06} & \heatcell{0}{47.37} & \heatcell{34}{82.09} & \heatcell{0}{20.39} & \heatcell{65}{35.63} & \heatcell{45}{11.44} & \heatcell{29}{29.49} \\ \midrule
\multirow{4}{*}{Qwen2.5-coder} 
& base & \heatcell{20}{4.30} & \heatcell{94}{58.91} & \heatcell{96}{83.14} & \heatcell{31}{27.91} & \heatcell{0}{29.55} & \heatcell{0}{8.61} & \heatcell{0}{27.13} \\
& ICL  & \heatcell{0}{2.03} & \heatcell{0}{45.90} & \heatcell{0}{78.69} & \heatcell{64}{36.07} & \heatcell{71}{43.50} & \heatcell{0}{8.64} & \heatcell{39}{31.21} \\
& RAG  & \heatcell{12}{3.40} & \heatcell{100}{59.80} & \heatcell{100}{83.33} & \heatcell{100}{45.10} & \heatcell{100}{49.28} & \heatcell{100}{15.10} & \heatcell{100}{37.57} \\
& COT  & \heatcell{100}{13.25} & \heatcell{37}{51.01} & \heatcell{33}{80.21} & \heatcell{0}{20.35} & \heatcell{97}{48.66} & \heatcell{30}{10.58} & \heatcell{23}{29.52} \\ \bottomrule
\end{tabular}
\end{table}

\subsection{RQ2: What's the best tuning strategy using SOLMs for automated logging?}

\textbf{Motivation.}
In the course of the fine-tuning operation, a diverse array of strategies has the potential to influence the efficacy of our tasks significantly. In order to systematically assess the capabilities of the SOLMs, we thoroughly investigate the following factors:

(RQ2.1) Which \textbf{PEFT technique} yields the optimal performance for automoated logging statement generation using SOLMs? While SOLMs are more compact than larger LLMs, fully fine-tuning them for specific downstream tasks like automated logging can still be computationally demanding and may risk overfitting on task-specific data. PEFT methodologies have emerged as a compelling solution, enabling adaptation by updating only a small fraction of the model's parameters or by adding a small set of new, trainable parameters. This significantly reduces computational costs and avoids catastrophic forgetting of the model's pre-trained knowledge. However, a diverse range of PEFT techniques exists, each employing different mechanisms to inject task-specific information into the SOLM. For automated logging, which involves understanding code context, identifying appropriate logging locations, and generating relevant log messages, it is unclear which PEFT strategy offers the optimal balance of performance and efficiency. Therefore, we conduct an evaluation of the performance of SOLMs fine-tuned with various prominent PEFT techniques to ascertain which techniques yield superior performance outcomes.

(RQ2.2) How does the \textbf{size of SOLMs} impact the performance-resource trade-offs in automated logging? Although we focus on SOLMs, there is still considerable variation in size within this class. Larger models might capture more complex code patterns and lead to higher accuracy, but they could also incur greater computational costs during fine-tuning and inference. Therefore, evaluating the impact of model size is essential to understand the performance-resource trade-offs specific to automated logging.

(RQ2.3) Does the \textbf{instruct} variant of a SOLM outperform its \textbf{base} counterpart for automated logging? The distinction between using a `base' pre-trained model versus an `instruct' version of an SOLM presents another critical strategic choice. \added{The `instruct' variants of SOLMs have been pre-tuned by their creators on a wide array of tasks to enhance their ability to follow general user prompts. In contrast, `base' models are not tuned for instruction-following and represent the direct output of the pre-training phase. This presents a critical choice: Is it more effective to start with a model already skilled in following general instructions, or does the base model offer greater plasticity for specialization when fine-tuned on a narrow, specific task such as automated logging? Therefore, we aim to clarify which model mode serves as a better foundation for our task.}

\added{(RQ2.4) Which prompting strategy is most effective when combined with fine-tuned SOLMs? Our prior findings established LoRA as the optimal fine-tuning method (RQ2.1) and identified RAG as the most effective prompting strategy for base models (RQ1). However, since fine-tuning fundamentally alters a model's behavior, we cannot assume that the best prompting strategy for a base model remains optimal for its fine-tuned counterpart. To build a methodologically sound basis for our final model evaluation in RQ3, we empirically determine which prompting strategy works most effectively in synergy with a LoRA-tuned model. This step is crucial to validate our choice of the definitive best-performing configuration.}

\textbf{Approach.}
To address RQ2.1, we selected the 7B parameter versions of all four SOLMs as our primary subjects for investigating the impact of different PEFT techniques. We systematically evaluated four prominent PEFT methods: Prefix Tuning, Prompt Tuning, LoRA, and QLoRA. To establish a comparative baseline, we also measured the performance using direct inference without any PEFT fine-tuning (referred to as `base'). For all experiments conducted under RQ2.1, including the baseline, we consistently utilized the RAG-enhanced prompt format that has been identified as effective during our experiment in RQ1.

For RQ2.2, focusing on the influence of model size, we selected the Qwen2.5-coder model series. This choice is driven by the public availability of multiple versions within the same model family, specifically those with 0.5B, 1.5B, 3B, 7B, and 14B parameters, enabling a controlled comparison. Based on the findings from RQ2.1 where LoRA demonstrated superior performance among the PEFT techniques, we exclusively employ LoRA for fine-tuning across these different model sizes. Furthermore, we evaluat each size using both the basic prompt (`base') and the RAG-enhanced prompt (`RAG'), in order to explore whether the effectiveness of RAG varies with model scale, particularly to assess the RAG capabilities of the smaller SOLMs.

To address RQ2.3, we compare the performance of `base' models against their `instruct' counterparts for automated logging. We select the Mistral 7B model, available in both base and instruct variants, for a controlled comparison, as it got the best perfomance in RQ2.1. Both model modes are fine-tuned using LoRA and employed the RAG-enhanced prompt, consistent with the result of RQ2.1 and RQ2.2. The fine-tuning process spanned five epochs, and performance are evaluated using metrics including PA, LA, AOD, PMR, F1, BLEU-4, and ROUGE-L. \added{To specifically assess the models’ adherence to instructions in this comparison, we also introduce a diagnostic metric: the Reject Rate. The Reject Rate is defined as the percentage of test set examples rejected by the fine-tuned model during inference due to misalignment with learned task-specific criteria; this quantifies the model’s selectivity, reflecting its ability to adhere to prompt instructions and generate valid logging statements.}

\added{To address RQ2.4, we conduct a controlled experiment applying all four prompting strategies to the LoRA-tuned versions of Qwen2.5-coder-7B. We evaluated the performance of four configurations (e.g., LoRA+ICL) on the test set. This comparative analysis allows us to isolate the impact of each prompting technique on an already specialized model and identify the most effective combination of fine-tuning and prompting for automated logging.}

\begin{table}[tbp]
\centering
\caption{Performance Comparison of PEFT Techniques for four 7B SOLMs in automated logging. The greener the better.}
\label{tab:2.1PEFT}
\setlength{\tabcolsep}{4pt} 
\begin{tabular}{@{}ll||c||cc|cc|cc}
\toprule
\multirow{2}{*}{Model} & \multirow{2}{*}{PEFT Techs} & Location & \multicolumn{2}{c}{Level} & \multicolumn{2}{c}{Variable} & \multicolumn{2}{c}{Text} \\ \cmidrule(l){3-9} 
& & PA & LA & AOD & PMR & F1 & BLEU-4 & ROUGE-L \\ \midrule
\multirow{5}{*}{LLaMA3} 
& base          & \heatcell{0}{21.01}   & \heatcell{0}{53.72}   & \heatcell{0}{82.09}    & \heatcell{31}{41.68}  & \heatcell{8}{46.28}    & \heatcell{28}{15.95} & \heatcell{20}{36.73} \\
& prefix tuning & \heatcell{0}{20.98}   & \heatcell{27}{57.46}  & \heatcell{26}{83.59}   & \heatcell{0}{37.94}   & \heatcell{0}{45.27}    & \heatcell{0}{14.42}  & \heatcell{0}{35.71}  \\
& prompt tuning & \heatcell{11}{25.01}  & \heatcell{100}{67.38} & \heatcell{100}{87.85}  & \heatcell{42}{43.01}  & \heatcell{39}{50.33}   & \heatcell{23}{15.66} & \heatcell{5}{35.96}  \\
& LoRA          & \heatcell{100}{56.84} & \heatcell{72}{63.56}  & \heatcell{92}{87.37}   & \heatcell{100}{50.15} & \heatcell{100}{58.22}  & \heatcell{100}{19.84} & \heatcell{100}{40.89} \\
& QLoRA         & \heatcell{68}{45.27}  & \heatcell{25}{57.18}  & \heatcell{26}{83.56}   & \heatcell{55}{44.66}  & \heatcell{47}{51.29}   & \heatcell{29}{15.97} & \heatcell{17}{36.60} \\ \midrule
\multirow{5}{*}{Mistral} 
& base          & \heatcell{0}{18.15}   & \heatcell{49}{65.14}  & \heatcell{33}{86.35}  & \heatcell{0}{37.25}   & \heatcell{0}{40.21}    & \heatcell{0}{14.36}  & \heatcell{0}{34.74}  \\
& prefix tuning & \heatcell{5}{20.35}   & \heatcell{28}{63.34}  & \heatcell{0}{85.20}   & \heatcell{11}{38.95}  & \heatcell{15}{43.07}   & \heatcell{18}{15.96} & \heatcell{18}{36.75} \\
& prompt tuning & \heatcell{30}{31.87}  & \heatcell{0}{60.92}   & \heatcell{6}{85.41}   & \heatcell{40}{43.68}  & \heatcell{71}{53.59}   & \heatcell{38}{17.81} & \heatcell{33}{38.36} \\
& LoRA          & \heatcell{100}{63.97} & \heatcell{100}{69.50} & \heatcell{99}{88.64}  & \heatcell{97}{52.79}  & \heatcell{94}{57.89}   & \heatcell{100}{23.40} & \heatcell{100}{45.73} \\
& QLoRA         & \heatcell{95}{61.90}  & \heatcell{97}{69.28}  & \heatcell{100}{88.66} & \heatcell{100}{53.31} & \heatcell{100}{58.97}  & \heatcell{89}{22.42} & \heatcell{91}{44.70} \\ \midrule
\multirow{5}{*}{CodeLlama} 
& base          & \heatcell{0}{8.16}    & \heatcell{0}{62.04}   & \heatcell{13}{86.43}  & \heatcell{0}{31.02}   & \heatcell{0}{39.01}    & \heatcell{0}{13.82}  & \heatcell{0}{34.36}  \\
& prefix tuning & \heatcell{16}{16.35}  & \heatcell{1}{62.12}   & \heatcell{0}{86.07}   & \heatcell{46}{41.14}  & \heatcell{17}{42.43}   & \heatcell{36}{17.19} & \heatcell{39}{38.48} \\
& prompt tuning & \heatcell{20}{18.15}  & \heatcell{6}{62.42}   & \heatcell{28}{86.81}  & \heatcell{49}{41.76}  & \heatcell{54}{49.91}   & \heatcell{49}{18.41} & \heatcell{51}{39.69} \\
& LoRA          & \heatcell{100}{59.01} & \heatcell{100}{68.57} & \heatcell{100}{88.74} & \heatcell{100}{52.77} & \heatcell{100}{59.10}  & \heatcell{100}{23.10} & \heatcell{100}{44.92} \\
& QLoRA         & \heatcell{100}{58.97} & \heatcell{95}{68.27}  & \heatcell{95}{88.62}  & \heatcell{96}{52.00}  & \heatcell{95}{58.12}   & \heatcell{91}{22.30} & \heatcell{96}{44.48} \\ \midrule
\multirow{5}{*}{Qwen2.5Coder} 
& base          & \heatcell{0}{3.40}    & \heatcell{0}{59.80}   & \heatcell{0}{83.33}    & \heatcell{12}{45.10}  & \heatcell{0}{49.28}    & \heatcell{0}{15.10}  & \heatcell{22}{37.57} \\
& prefix tuning & \heatcell{37}{25.27}  & \heatcell{66}{65.48}  & \heatcell{81}{87.80}   & \heatcell{0}{44.14}   & \heatcell{69}{56.36}   & \heatcell{2}{15.23}  & \heatcell{0}{35.47}  \\
& prompt tuning & \heatcell{45}{30.04}  & \heatcell{91}{67.63}  & \heatcell{85}{88.01}   & \heatcell{24}{46.12}  & \heatcell{52}{54.63}   & \heatcell{47}{18.80} & \heatcell{34}{38.74} \\
& LoRA          & \heatcell{100}{62.40} & \heatcell{99}{68.46}  & \heatcell{100}{88.82}  & \heatcell{100}{52.24} & \heatcell{95}{58.98}   & \heatcell{100}{23.04} & \heatcell{100}{44.96} \\
& QLoRA         & \heatcell{96}{60.21}  & \heatcell{93}{67.87}  & \heatcell{98}{88.70}   & \heatcell{92}{51.55}  & \heatcell{100}{59.47}  & \heatcell{93}{22.52} & \heatcell{90}{44.03} \\ \bottomrule
\end{tabular}
\end{table}

\textbf{Results.}
(RQ2.1) Table~\ref{tab:2.1PEFT} shows the performance comparison of PEFT techniques for SOLMs, where we observe that all evaluated PEFT methods consistently improve performance over the baseline across all SOLMs and most metrics. For instance, looking at the prediction accuracy, QLoRA fine-tuning increased PA from 13.32 to 45.27 for LLaMA3, from 12.85 to 61.90 for Mistral, from 6.99 to 58.97 for CodeLlama, and from 4.30 to 60.21 for Qwen2.5Coder. Similar substantial gains are observed across other metrics like F1 score for variable prediction and BLEU-4/ROUGE-L for statement generation, indicating that fine-tuning with parameter-efficient techniques, is crucial for adapting SOLMs to this specific task. 

Furthermore, LoRA emerges as the most effective PEFT methodology, achieving the highest scores for the majority of metrics across all four models. For example, with LoRA, LLaMA3 gets the top PA (56.84), PMR (50.15), F1 (58.22), BLEU-4 (19.84), and ROUGE-L (40.89). Similarly, LoRA leads to the best PA (63.97), LA (69.50), BLEU-4 (23.40), and ROUGE-L (45.73) for Mistral. CodeLlama shows LoRA as the top performer across all metrics. For Qwen2.5Coder, LoRA is also dominant, securing the best PA (62.40), LA (68.46), AOD (88.82), PMR (52.24), BLEU-4 (23.04), and ROUGE-L (44.96). Moreover, QLoRA is competitive to LoRA, achieving the second-best results or even surpassing LoRA in a few specific instances (e.g., F1 for Qwen2.5Coder, PMR, and F1 for Mistral). Prompt Tuning shows some strength, particularly for `level' prediction with LLaMA3, while Prefix Tuning, though an improvement over the baseline, is generally outperformed by LoRA, QLoRA, and Prompt Tuning.

\begin{figure}[t]
\centering
\begin{subfigure}{0.32\textwidth}
\centering
\includegraphics[width=\linewidth]{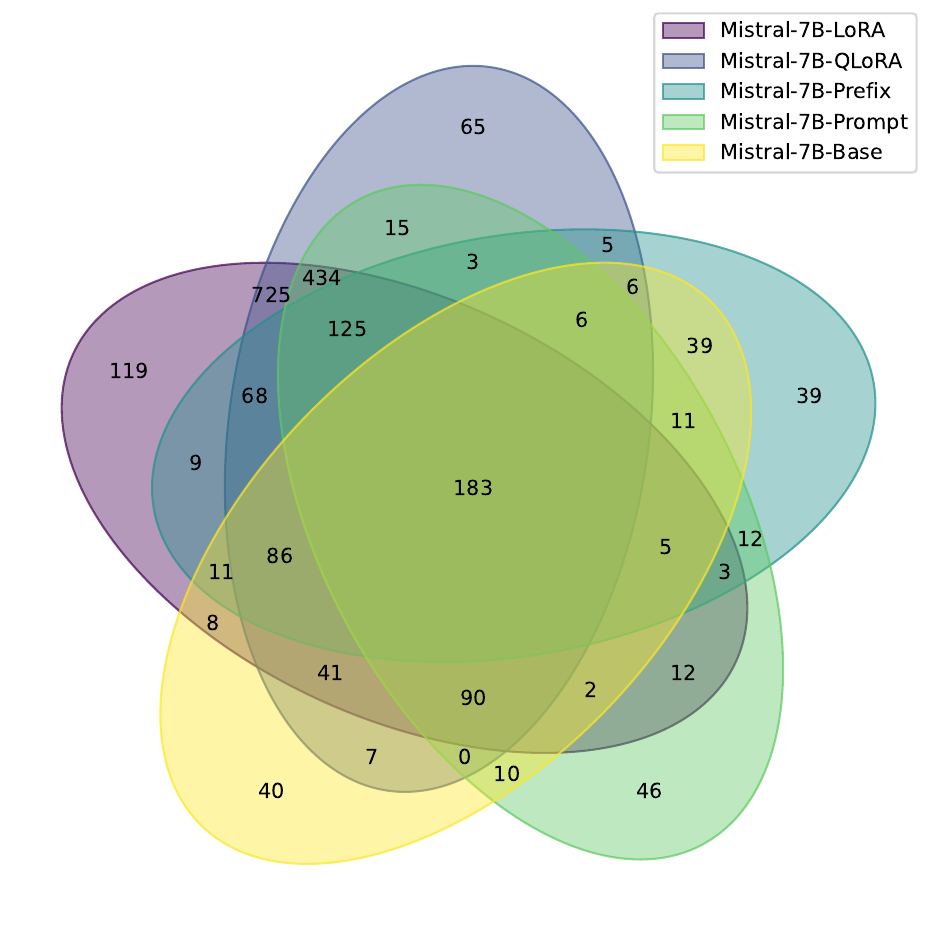}
\caption{Mistral-7B with PEFT Variations}
\label{fig:venn21}
\end{subfigure}
\hfill
\begin{subfigure}{0.32\textwidth}
\centering
\includegraphics[width=\linewidth]{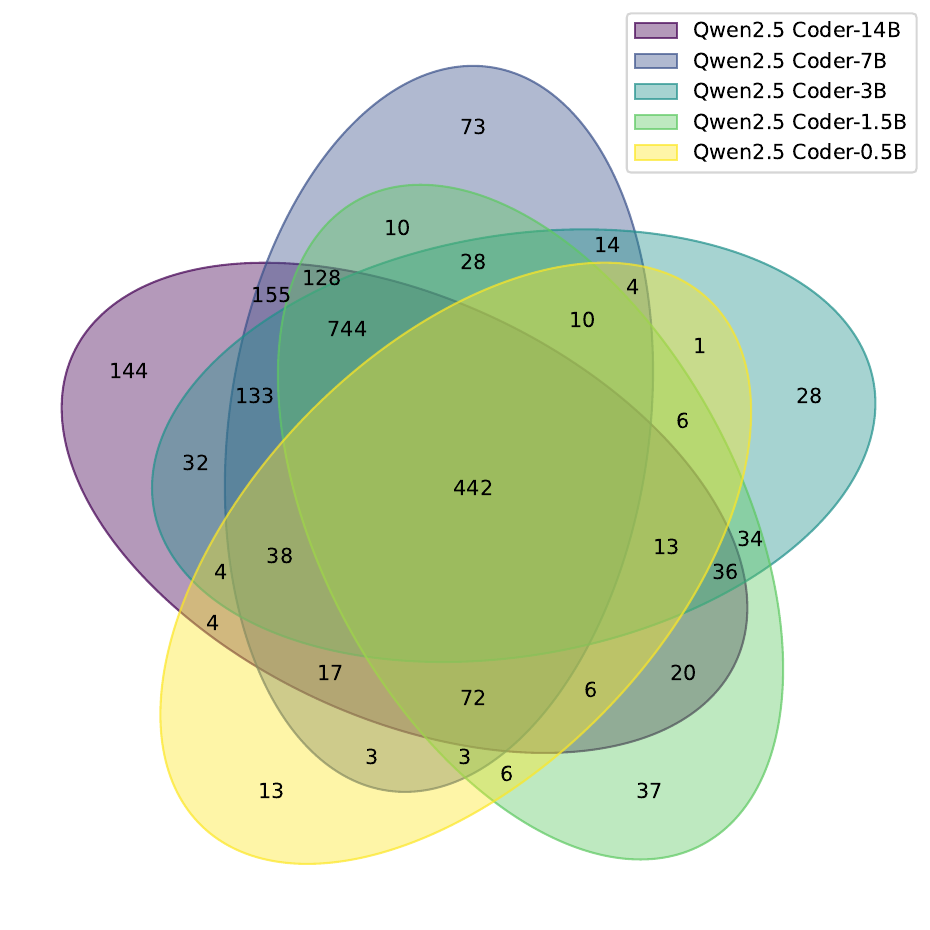}
\caption{Qwen2.5-coder of Different Sizes}
\label{fig:venn22}
\end{subfigure}
\hfill
\begin{subfigure}{0.32\textwidth}
\centering
\includegraphics[width=\linewidth]{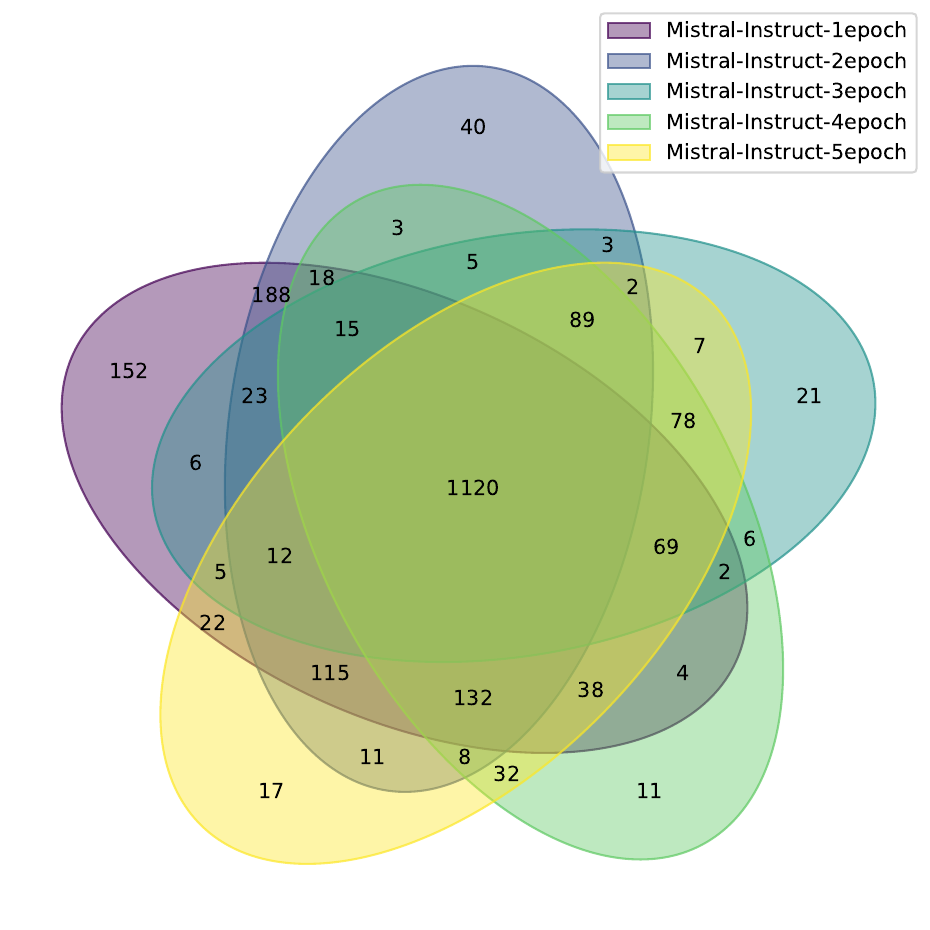}
\caption{Mistral-Instruct-7B Across Epochs}
\label{fig:venn23}
\end{subfigure}
\caption{Overlap of Correctly Logging Statement Placement Across Different SOLM Configurations.}
\label{fig:three_venn}
\end{figure}

\added{Additionally, the venn diagram in Figure~\ref{fig:venn21} illustrates the overlap of correctly predicted logging statement placements across different PEFT variations for Mistral-7B. The largest overlap (183) is observed in the central region, indicating a core set of logging statements whose placement were successfully identified by all PEFT methods .} LoRA and QLoRA show significant individual contributions (125 and 86, respectively), suggesting their effectiveness in identifying unique logging placements, while the base method contributes the least (40), highlighting the improvement brought by PEFT techniques.

\begin{tcolorbox}[boxsep=1pt,left=2pt,right=2pt,top=3pt,bottom=2pt,width=\columnwidth,colback=white!95!black,boxrule=0pt, colbacktitle=white!30!black,toptitle=2pt,bottomtitle=1pt,opacitybacktitle=0.4, sharp corners]

\textbf{Findings 3}: Fine-tuning with PEFT techniques significantly enhances SOLMs performance for automated logging, with LoRA demonstrating the most consistent and superior results across the evaluated models and metrics.

\end{tcolorbox}

\begin{table}[t]
\centering
\caption{Impact of Model Size on Performance and Resource Usage for automated logging. The greener, the better.}
\label{tab:2.2modelsize}
\setlength{\tabcolsep}{5pt} 
\scriptsize
\begin{tabular}{rrrrrrrrrrr}
\toprule
\shortstack[r]{Model \\ params} & \shortstack[r]{Trainable \\ params} & \shortstack[r]{Training time \\ (s/epoch)} & \shortstack[r]{Inference time\\ (s/prompt)} & PA & LA & AOD & PMR & F1 & BLEU-4 & ROUGE-L \\ \midrule
0.5B & \textasciitilde281M  & \heatcell{100}{2438.2749} & \heatcell{100}{0.0959} & \heatcell{0}{21.38}   & \heatcell{9}{61.53}  & \heatcell{31}{86.78}  & \heatcell{0}{41.74}   & \heatcell{0}{56.59}   & \heatcell{0}{18.95}  & \heatcell{0}{39.55}  \\
1.5B & \textasciitilde485M  & \heatcell{90}{4395.2893}  & \heatcell{77}{0.1866}  & \heatcell{71}{53.11}  & \heatcell{0}{60.69}   & \heatcell{0}{85.70}   & \heatcell{67}{50.34}  & \heatcell{54}{58.39}  & \heatcell{22}{20.35} & \heatcell{15}{40.69} \\
3B   & \textasciitilde652M  & \heatcell{73}{7694.3684}  & \heatcell{64}{0.2375}  & \heatcell{69}{52.18}  & \heatcell{84}{68.41}  & \heatcell{90}{88.77}  & \heatcell{58}{49.20}  & \heatcell{10}{56.94}  & \heatcell{52}{22.23} & \heatcell{47}{43.18} \\
7B   & \textasciitilde1139M & \heatcell{44}{13258.3642} & \heatcell{55}{0.2710}  & \heatcell{91}{62.40}  & \heatcell{84}{68.46}  & \heatcell{91}{88.82}  & \heatcell{82}{52.24}  & \heatcell{72}{58.98}  & \heatcell{65}{23.04} & \heatcell{71}{44.96} \\
14B  & \textasciitilde1625M & \heatcell{0}{21780.5427}  & \heatcell{0}{0.4854}   & \heatcell{100}{66.20} & \heatcell{100}{69.92} & \heatcell{100}{89.36} & \heatcell{100}{54.53} & \heatcell{100}{59.93} & \heatcell{100}{25.20} & \heatcell{100}{47.22} \\ \bottomrule
\end{tabular}
\end{table}

(RQ2.2) Table~\ref{tab:2.2modelsize} reveals that increasing model size leads to improved performance in automated logging, particularly for models 3B and larger. Across almost all metrics, there is a discernible improvement as the model parameter count increases from 0.5B to 14B. For example, prediction accuracy (PA) improves from 21.38 (0.5B) to 66.20 (14B), and ROUGE-L scores for text generation increase from 39.55 (0.5B) to 47.22 (14B). The 14B model consistently outperforms all smaller variants, and the 7B model also shows strong performance.

Models smaller than 3B (i.e., 0.5B and 1.5B) exhibit less stable performance scaling. While the 0.5B model is generally the weakest, the progression to the 1.5B and then to the 3B model is not uniformly positive across all metrics. For instance, the 1.5B model shows a slight decrease in LA (60.69 vs 61.53 for 0.5B) and AOD (85.70 vs 86.78 for 0.5B). Furthermore, when moving from 1.5B to 3B, there are slight dips in PA (52.18 vs 53.11), PMR (49.20 vs 50.34), and F1 (56.94 vs 58.39). This suggests that while larger models generally perform better, the performance gains for models below 3B parameters might be less consistent or predictable for this specific task and fine-tuning approach.

Larger models come with increased resource requirements. As model size increases, the training time per epoch, and inference time per prompt also escalate substantially. For instance, training time per epoch rises from approximately 2438 seconds for the 0.5B model to 21780 seconds for the 14B model. Similarly, inference time per prompt increases from 0.0959 seconds (0.5B) to 0.4854 seconds (14B).

\added{The venn diagram in Figure~\ref{fig:venn22} depicts the overlap of correctly predicted logging statement placements across different sizes of the Qwen2.5-coder model (0.5B, 1.5B, 3B, 7B, 14B). The central overlap (442) represents logging statements whose placements were successfully identified across all model sizes, with the 14B model contributing the most unique placements (152), followed by 7B (73).} Smaller models (0.5B and 1.5B) show limited unique contributions (21 and 6, respectively), reinforcing the finding that models with 3B+ parameters perform better for this task.

\begin{tcolorbox}[boxsep=1pt,left=2pt,right=2pt,top=3pt,bottom=2pt,width=\columnwidth,colback=white!95!black,boxrule=0pt, colbacktitle=white!30!black,toptitle=2pt,bottomtitle=1pt,opacitybacktitle=0.4, sharp corners]

\textbf{Findings 4}: For models with 3B+ parameters, performance in generating logging statements improves with more parameters, but increases computational costs, indicating a performance-resource trade-off. Models under 3B show inconsistent scaling, suggesting a minimum capacity may needed for this task.

\end{tcolorbox}

\begin{figure}[tbp]
\centering

\hspace*{\fill}
\begin{subfigure}{0.48\textwidth}
\centering
\includegraphics[width=\linewidth]{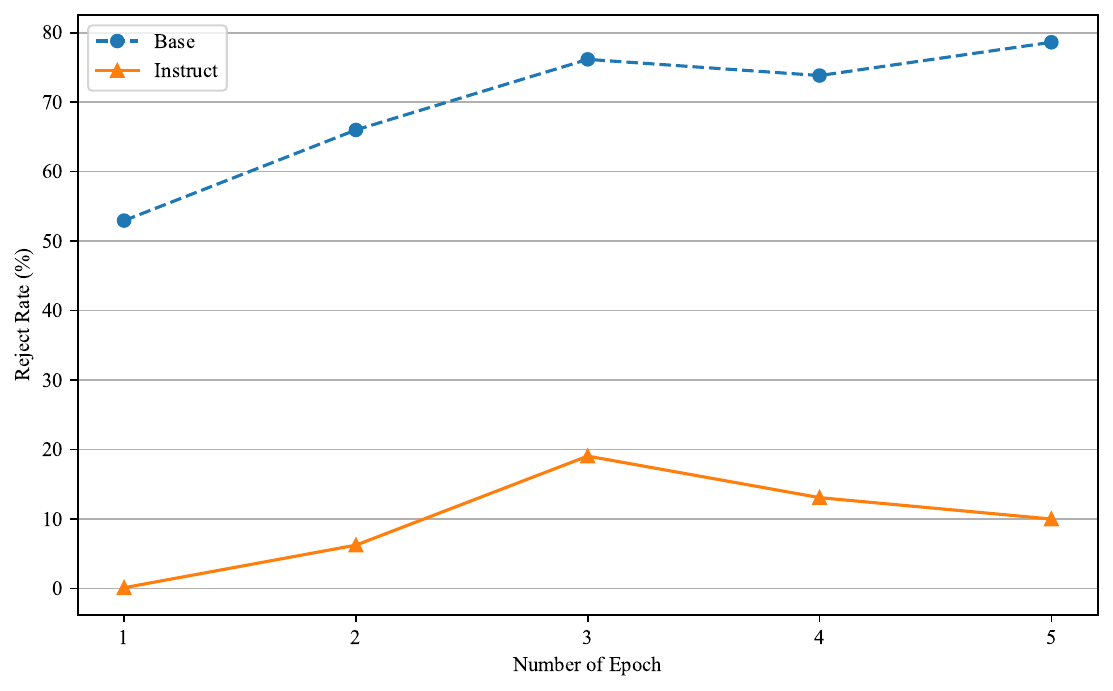}
\caption{Base vs. Instruct on Reject Rate}
\label{fig:rq23sub1}
\end{subfigure}\hfill
\begin{subfigure}{0.48\textwidth}
\centering
\includegraphics[width=\linewidth]{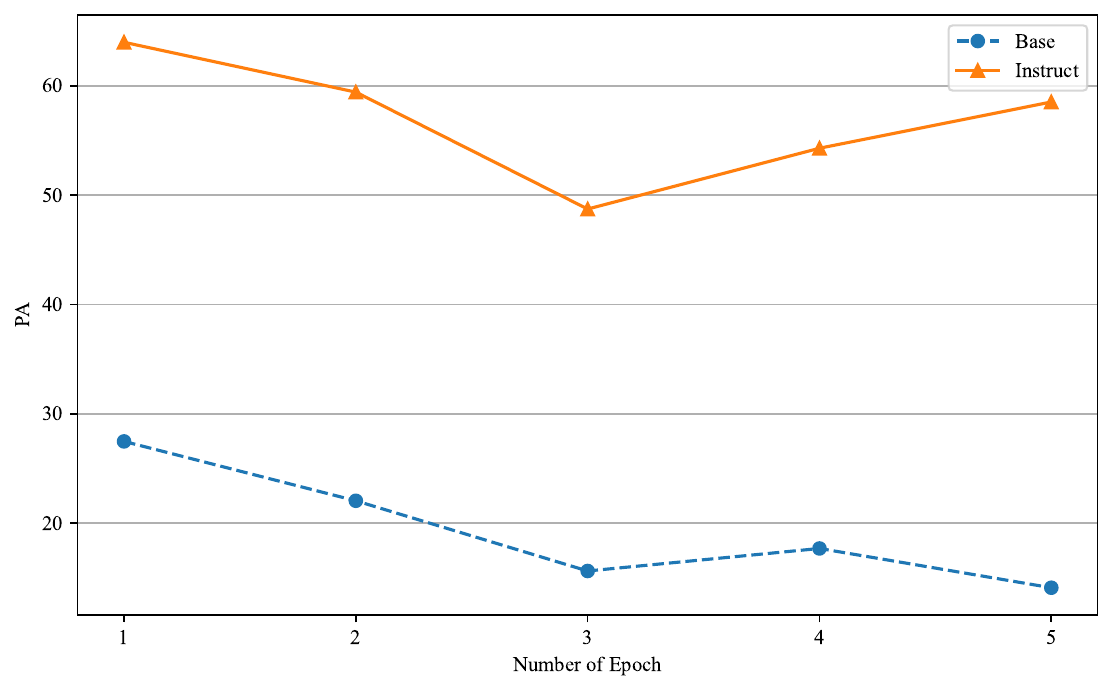}
\caption{Base vs. Instruct on PA}
\label{fig:rq23sub2}
\end{subfigure}
\hspace*{\fill}

\vspace{0.5cm}

\hspace*{\fill}
\begin{subfigure}{0.48\textwidth}
\centering
\includegraphics[width=\linewidth]{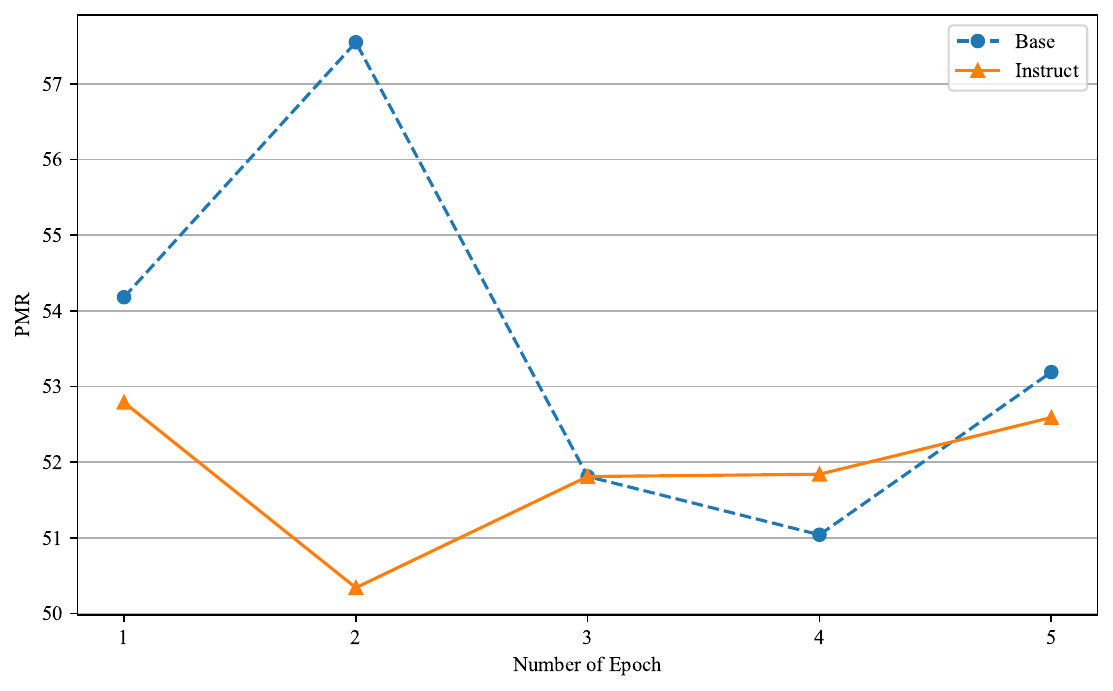}
\caption{Base vs. Instruct on PMR}
\label{fig:rq23sub3}
\end{subfigure}\hfill
\begin{subfigure}{0.48\textwidth}
\centering
\includegraphics[width=\linewidth]{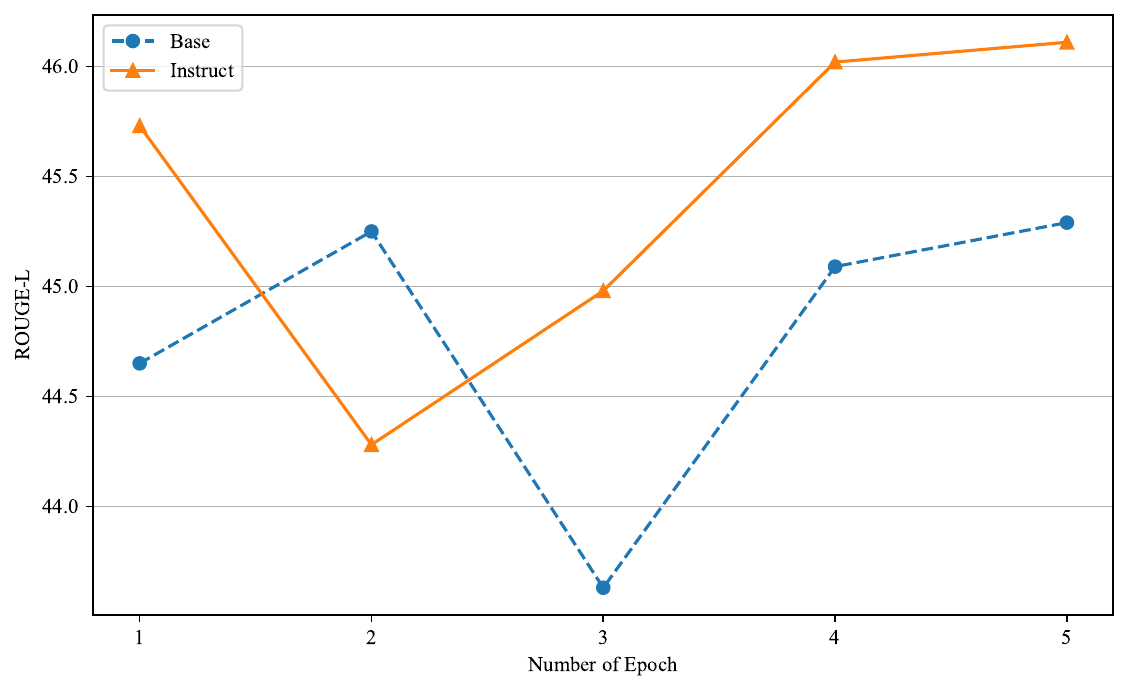} 
\caption{Base vs. Instruct on ROUGE-L}
\label{fig:rq23sub4}
\end{subfigure}\hfill
\hspace*{\fill}

\caption{Performance Comparison of Base and Instruct Mistral-7B Models Across Five Epochs.}
\label{fig:rq23}
\end{figure}

(RQ2.3) Figure~\ref{fig:rq23} illustrates the performance comparison between the base and instruct variants of the Mistral-7B model across five epochs for automated logging. The instruct model consistently outperforms the base model across most metrics, including Reject Rate, PA, and ROUGE-L. For instance, the instruct model achieves a lower Reject Rate, indicating better adherence to task-specific criteria and fewer invalid outputs (Figure~\ref{fig:rq23sub1}). Similarly, the instruct model demonstrates higher PA (Figure~\ref{fig:rq23sub2}), reflecting superior accuracy in predicting logging statement placement.
The ROUGE-L scores (Figure~\ref{fig:rq23sub4}) further confirm that the instruct model generates logging statements with greater textual similarity to the ground truth. These trends are evident from the first epoch and persist through the fifth, suggesting that the instruct model’s pre-training for instruction-following enhances its ability to adapt to our logging task.

\begin{tcolorbox}[boxsep=1pt,left=2pt,right=2pt,top=3pt,bottom=2pt,width=\columnwidth,colback=white!95!black,boxrule=0pt, colbacktitle=white!30!black,toptitle=2pt,bottomtitle=1pt,opacitybacktitle=0.4, sharp corners]

\textbf{Findings 5}: The instruct variant of SOLM model outperforms its base counterpart in automated logging, benefiting from its instruction-tuned foundation, which enhances task adherence and output quality.

\end{tcolorbox}

In addition, the performance trends across epochs reveal that excessive fine-tuning can lead to diminishing returns. For both models, the Reject Rate and PA peak at the first epoch, where the Reject Rate is minimized, and the number of correctly predicted logging statement placements is maximized (Figure~\ref{fig:rq23sub1} and Figure~\ref{fig:rq23sub2}). Beyond the first epoch, both metrics show a slight decline or stabilization, with the Reject Rate marginally increasing and PA slightly decreasing by the fifth epoch. This trend suggests that additional fine-tuning may lead to overfitting, causing the models to become overly specialized to the training data and potentially losing some generalizability. For the instruct model, this could also imply a partial erosion of its pre-trained instruction-following robustness.

\added{The venn diagram in Figure~\ref{fig:venn23} further illustrates the overlap of correctly predicted logging statement placements across epochs for the Mistral-7B-Instruct model. The central overlap (1120) indicates a stable core of accurately placed statements that were consistently identified across all five epochs, with the first epoch contributing the most unique correctly predicted placements (40).} The decline in unique contributions from later epochs (e.g., 3 for the fifth epoch) supports the observation of diminishing returns and potential overfitting with prolonged fine-tuning.

\begin{tcolorbox}[boxsep=1pt,left=2pt,right=2pt,top=3pt,bottom=2pt,width=\columnwidth,colback=white!95!black,boxrule=0pt, colbacktitle=white!30!black,toptitle=2pt,bottomtitle=1pt,opacitybacktitle=0.4, sharp corners]

\textbf{Findings 6}: Both base and instruct models achieve optimal performance at the first epoch for Reject Rate and PA, with prolonged fine-tuning leading to slight performance declines due to overfitting, indicating that excessive fine-tuning may compromise pre-trained capabilities.

\end{tcolorbox}

\begin{table}[tbp]
\centering
\caption{\added{Performance Comparison of Different Prompting Strategies Combined with LoRA.}}
\label{tab:rq24}
\begin{tabular}{lccccccc}
\toprule
\multicolumn{1}{c}{\multirow{2}{*}{Strategy}} & Location & \multicolumn{2}{c}{Level} & \multicolumn{2}{c}{Variable} & \multicolumn{2}{c}{Text} \\
\multicolumn{1}{c}{} & PA & LA & AOD & PMR & F1 & BLEU-4 & ROUGE-L \\
\midrule
LoRA+base & 57.81 & 68.30 & 88.01 & 51.96 & 58.33 & 20.63 & 42.69 \\
LoRA+ICL & 60.77 & 66.08 & 87.59 & 51.45 & 57.75 & 21.39 & 44.06 \\
LoRA+RAG & \textbf{62.40} & \textbf{68.46} & \textbf{88.82} & \textbf{52.24} & \textbf{58.98} & \textbf{23.04} & \textbf{44.96} \\
LoRA+COT & 58.81 & 68.09 & 88.73 & 50.91 & 57.39 & 21.30 & 43.36 \\
\bottomrule
\end{tabular}
\end{table}

\added{(RQ2.4) Table~\ref{tab:rq24} presents the performance comparison of different prompting strategies when applied to the LoRA-tuned Qwen2.5-coder-7B model. The results clearly indicate that the combination of LoRA and RAG yields the most effective performance across all evaluated metrics. This configuration achieved the highest scores in position accuracy (i.e., PA: 62.40\%), level accuracy (i.e., LA: 68.46\%), variable prediction (i.e., F1: 58.98\%), and text generation quality (i.e., BLEU-4: 23.04, ROUGE-L: 44.96\%). }

\added{While other prompting strategies also benefited from the LoRA fine-tuning, their performance was consistently surpassed by the RAG-enhanced approach. For instance, LoRA+ICL, the second-best strategy, achieved a PA of 60.77\%, which is 1.63 percentage points lower than LoRA+RAG. This outcome validates the hypothesis that the benefits of LoRA and RAG are synergistic. Fine-tuning with LoRA adapts the model to the specific structural and syntactic nuances of the automated logging task, while RAG provides critical, instance-specific context at inference time. This combination allows the specialized model to leverage the most relevant examples, leading to a significant performance boost. Therefore, we confirm that LoRA fine-tuning combined with RAG prompting is the optimal strategy for this task.}

\begin{tcolorbox}[boxsep=1pt,left=2pt,right=2pt,top=3pt,bottom=2pt,width=\columnwidth,colback=white!95!black,boxrule=0pt, colbacktitle=white!30!black,toptitle=2pt,bottomtitle=1pt,opacitybacktitle=0.4, sharp corners]

\added{\textbf{Findings 7:} The RAG prompting strategy is most effective when combined with a LoRA-tuned model, outperforming other prompting techniques across all metrics. This confirms a synergistic effect between task-specific fine-tuning and context-aware prompting.}

\end{tcolorbox}
\subsection{RQ3: How effectively do SOLMs compare to existing methods and LLM baselines in automated logging?}

\textbf{Motivation.}
Having established optimal strategies for employing SOLMs in addressing the preceding RQs, this study aims to investigate the performance of SOLMs in automated logging compared to existing methods and the direct application of LLMs.

\begin{itemize}
    \item (RQ3.1) How effectively do these methods determine appropriate logging locations?
    \item (RQ3.2) What is the quality of the logging statements generated by these methods?
    \item (RQ3.3) When evaluated by an LLM acting as a judge, how does the overall quality of the logging statements produced by these methods compare?
\end{itemize}

\textbf{Approach.}
To address RQ3, we evaluate the performance of SOLMs in automated logging against existing method (i.e., LANCE, LEONID, Unilog, and Fastlog) and LLMs (i.e., Claude3.7sonnet, Deepseek-coder-v3, GPT4o, and LLaMA3.1-405B). We assess the loging location accuracy (PA), the statement quality (LA, AOD, PMR, F1, BLEU-4, and ROUGE-L), and overall quality via our LLM judges. Target SOLMs (LLaMA-8B, Mistral-7B, CodeLlama-13B, Qwen2.5-coder-14B) are fine-tuned with LoRA and RAG, while LLMs were tested in base, ICL, RAG, and COT configurations. To ensure a comprehensive evaluation of SOLMs’ capabilities, we select the largest parameter models available within the SOLM definition (i.e., open-source models with fewer than 14B parameters). This choice maximizes the potential performance of SOLMs, allowing us to showcase their optimal effectiveness in automated logging.

\begin{table}[tbp]
\centering
\caption{Performance Comparison of Automated Logging Approaches Across All Metrics. The greener, the better.}
\small
\label{tab:RQ3_merged}
\setlength{\tabcolsep}{5pt} 
\begin{tabular}{l||c||cc|cc|cc}
\toprule
& Location & \multicolumn{2}{c|}{Level} & \multicolumn{2}{c|}{Variable} & \multicolumn{2}{c}{Text} \\
Model & PA & LA & LOD & PMR & F1 & BLEU-4 & ROUGE-L \\
\midrule
\rowcolor{grey}\multicolumn{8}{c}{Existing Approach} \\
\midrule
LANCE & \heatcell{3}{44.67} & \heatcell{0}{48.23} & \heatcell{0}{80.60} & \heatcell{0}{26.84} & \heatcell{0}{48.33} & \heatcell{0}{11.21} & \heatcell{0}{29.38} \\
LEONID & \heatcell{26}{46.74} & \heatcell{7}{49.67} & \heatcell{16}{81.88} & \heatcell{5}{28.04} & \heatcell{12}{49.65} & \heatcell{13}{12.63} & \heatcell{13}{31.33} \\
Unilog & \heatcell{76}{51.27} & \heatcell{41}{56.34} & \heatcell{50}{84.58} & \heatcell{31}{34.19} & \heatcell{20}{50.46} & \heatcell{25}{14.05} & \heatcell{35}{34.68} \\
Fastlog & \heatcell{100}{53.40} & \heatcell{56}{59.26} & \heatcell{58}{85.23} & \heatcell{48}{38.22} & \heatcell{27}{51.24} & \heatcell{18}{13.28} & \heatcell{21}{32.54} \\
\added{SCLogger} & \added{\heatcell{0}{44.39}} & \added{\heatcell{100}{68.02}} & \added{\heatcell{100}{88.56}} & \added{\heatcell{100}{50.54}} & \added{\heatcell{100}{59.17}} & \added{\heatcell{100}{22.48}} & \added{\heatcell{100}{44.60}} \\
\midrule
\rowcolor{grey}\multicolumn{8}{c}{Proprietary Large Language Model} \\
\midrule
Claude3.7sonnet-base & \heatcell{63}{47.02} & \heatcell{84}{66.22} & \heatcell{95}{87.83} & \heatcell{67}{46.25} & \heatcell{77}{55.14} & \heatcell{65}{17.32} & \heatcell{83}{39.85} \\
Claude3.7sonnet-ICL & \heatcell{94}{62.80} & \heatcell{55}{62.46} & \heatcell{70}{86.64} & \heatcell{58}{44.70} & \heatcell{75}{55.06} & \heatcell{29}{14.93} & \heatcell{49}{36.84} \\
Claude3.7sonnet-RAG & \heatcell{100}{65.90} & \heatcell{81}{65.89} & \heatcell{98}{88.01} & \heatcell{65}{45.78} & \heatcell{100}{56.49} & \heatcell{67}{17.39} & \heatcell{81}{39.64} \\
Claude3.7sonnet-COT & \heatcell{63}{47.32} & \heatcell{72}{64.60} & \heatcell{97}{87.95} & \heatcell{2}{35.33} & \heatcell{84}{55.58} & \heatcell{32}{15.13} & \heatcell{49}{36.87} \\
Deepseek-coder-v3-base & \heatcell{76}{53.65} & \heatcell{96}{67.85} & \heatcell{99}{88.04} & \heatcell{79}{48.23} & \heatcell{77}{55.14} & \heatcell{67}{17.42} & \heatcell{67}{38.40} \\
Deepseek-coder-v3-ICL & \heatcell{94}{62.64} & \heatcell{68}{64.06} & \heatcell{75}{86.89} & \heatcell{72}{46.94} & \heatcell{51}{53.76} & \heatcell{45}{15.98} & \heatcell{51}{37.07} \\
Deepseek-coder-v3-RAG & \heatcell{99}{65.20} & \heatcell{100}{68.39} & \heatcell{100}{88.08} & \heatcell{86}{49.28} & \heatcell{90}{55.93} & \heatcell{91}{18.97} & \heatcell{100}{41.32} \\
Deepseek-coder-v3-COT & \heatcell{67}{49.05} & \heatcell{75}{65.11} & \heatcell{84}{87.31} & \heatcell{0}{34.96} & \heatcell{94}{56.12} & \heatcell{20}{14.31} & \heatcell{31}{35.32} \\
GPT4o-base & \heatcell{25}{27.91} & \heatcell{64}{63.60} & \heatcell{75}{86.89} & \heatcell{76}{47.73} & \heatcell{30}{52.62} & \heatcell{40}{15.66} & \heatcell{50}{36.95} \\
GPT4o-ICL & \heatcell{73}{52.45} & \heatcell{10}{56.44} & \heatcell{28}{84.63} & \heatcell{57}{44.51} & \heatcell{37}{52.96} & \heatcell{0}{13.02} & \heatcell{0}{32.56} \\
GPT4o-RAG & \heatcell{80}{55.78} & \heatcell{62}{63.28} & \heatcell{72}{86.75} & \heatcell{67}{46.27} & \heatcell{59}{54.20} & \heatcell{40}{15.66} & \heatcell{50}{36.94} \\
GPT4o-COT & \heatcell{19}{24.84} & \heatcell{58}{62.73} & \heatcell{76}{86.91} & \heatcell{21}{38.47} & \heatcell{96}{56.22} & \heatcell{31}{15.03} & \heatcell{44}{36.46} \\
LLaMA3.1-405B-base & \heatcell{62}{46.45} & \heatcell{66}{63.80} & \heatcell{77}{86.96} & \heatcell{100}{51.76} & \heatcell{68}{54.70} & \heatcell{84}{18.50} & \heatcell{78}{39.37} \\
LLaMA3.1-405B-ICL & \heatcell{72}{51.62} & \heatcell{0}{55.10} & \heatcell{0}{83.25} & \heatcell{69}{46.45} & \heatcell{19}{52.01} & \heatcell{30}{15.01} & \heatcell{36}{35.71} \\
LLaMA3.1-405B-RAG & \heatcell{79}{55.04} & \heatcell{64}{63.64} & \heatcell{76}{86.94} & \heatcell{96}{51.06} & \heatcell{92}{56.04} & \heatcell{100}{19.55} & \heatcell{96}{40.94} \\
LLaMA3.1-405B-COT & \heatcell{0}{15.25} & \heatcell{44}{60.92} & \heatcell{41}{85.23} & \heatcell{28}{39.74} & \heatcell{0}{50.94} & \heatcell{79}{18.22} & \heatcell{59}{37.71} \\
\midrule
\rowcolor{grey}\multicolumn{8}{c}{Fine-tuned Small Open-source Language Model} \\
\midrule
LLaMA3-8B-RAG-LoRA & \heatcell{0}{56.84} & \heatcell{0}{63.56} & \heatcell{9}{87.37} & \heatcell{0}{50.15} & \heatcell{16}{58.22} & \heatcell{0}{19.37} & \heatcell{7}{42.03} \\
Mistral-7B-RAG-LoRA & \heatcell{76}{63.97} & \heatcell{93}{69.50} & \heatcell{67}{88.64} & \heatcell{60}{52.79} & \heatcell{0}{57.89} & \heatcell{25}{20.62} & \heatcell{0}{41.70} \\
CodeLlama-13B-RAG-LoRA & \heatcell{72}{63.57} & \heatcell{14}{64.43} & \heatcell{0}{87.17} & \heatcell{86}{53.90} & \heatcell{34}{58.59} & \heatcell{100}{24.39} & \heatcell{100}{46.91} \\
Qwen2.5-coder-14B-RAG-LoRA & \heatcell{100}{66.20} & \heatcell{100}{69.92} & \heatcell{100}{89.36} & \heatcell{100}{54.53} & \heatcell{100}{59.93} & \heatcell{93}{24.05} & \heatcell{92}{46.51} \\
\bottomrule
\end{tabular}
\end{table}

\textbf{Result.}
(RQ3.1) Table~\ref{tab:RQ3_merged} shows the performance comparison of automated logging approaches across all metrics. The table shows that Qwen2.5-coder-14B-RAG-LoRA achieves the highest PA at 66.20\%, outperforming all other models, including LLMs like Claude3.7sonnet-RAG (65.90\%), Deepseek-coder-v3-RAG (65.20\%), as well as existing methods like Fastlog (53.40\%). Among SOLMs, Mistral-7B-RAG-LoRA (63.97\%) and CodeLlama-13B-RAG-LoRA (63.57\%) also surpass most LLMs and all existing methods, indicating that fine-tuning with LoRA and RAG significantly enhances the ability of SOLMs to identify logging locations in various configuration settings.

(RQ3.2)
\added{For statement quality, fine-tuned SOLMs demonstrate exceptional performance, outperforming both proprietary LLMs and all existing methods, including the strong SCLogger baseline. Qwen2.5-coder-14B-RAG-LORA leads on several key metrics, including the highest LA (69.92\%), AOD (89.36\%), and F1 score (59.93\%). Notably, these results surpass not only the best-performing LLM, deepseek-coder-v3-RAG (e.g., 68.39\% LA, 18.97\% BLEU-4), but also the highly competitive SCLogger (e.g., 68.02\% LA, 59.17\% F1).}

\added{Furthermore, CodeLlama-13B-RAG-LORA excels in text generation quality, achieving the highest BLEU-4 (24.39) and ROUGE-L (46.91) scores. This performance is superior to that of SCLogger, which itself sets a high bar with a BLEU-4 of 22.48 and a ROUGE-L of 44.60. The ability of fine-tuned SOLMs to outperform such a strong, specialized tool underscores the effectiveness of SOLMs. This suggests that targeted optimization enables SOLMs to generate more accurate, relevant, and contextually appropriate logging statements than both proprietary LLMs and state-of-the-art specialized tools.}

\begin{table}[tbp]
\centering
\small
\caption{\added{Correct Predictions Considering the Three-dimensional Challenges of Automated Logging.}}
\label{tab:threedimension}
\begin{tabular}{lccccc}
\toprule
Position & \XSolidBrush & \Checkmark & \Checkmark & \Checkmark & \Checkmark \\
Level & --- & \XSolidBrush & \Checkmark & \XSolidBrush & \Checkmark \\
Variable & --- & \XSolidBrush & \XSolidBrush & \Checkmark & \Checkmark \\
\midrule
\rowcolor{grey}\multicolumn{6}{c}{Proprietary Large Language Model} \\
\midrule
Claude3.7sonnet-RAG & 34.10 & 13.22 & 22.51 & 9.26 & 20.91 \\
GPT4o-RAG & 44.22 & 12.22 & 17.75 & 8.26 & 17.55 \\
Deepseek-coder-v3-RAG & 34.80 & 12.99 & 20.08 & 7.63 & 24.51 \\
LLaMA3.1-405B-RAG & 44.96 & 10.99 & 15.95 & 9.02 & 19.08 \\
\midrule
\rowcolor{grey}\multicolumn{6}{c}{Fine-tuned Small Open-source Language Models} \\
\midrule
LLaMA3-8B-RAG-LoRA & 43.16 & 11.69 & 16.65 & 9.02 & 19.48 \\
Mistral-7B-RAG-LoRA & 36.03 & 10.92 & 19.28 & 8.59 & 25.17 \\
CodeLlama-13B-RAG-LoRA & 36.43 & 11.72 & 17.58 & 10.89 & 25.38 \\
Qwen2.5-coder-14B-RAG-LoRA & 33.80 & 11.56 & 18.55 & 8.36 & 27.74 \\
\bottomrule
\end{tabular}
\end{table}

\added{To further address the relationship between metrics, we analyze the proportion of logging statements that are simultaneously correctly generated across the key dimensions of location, level, and variables. Table~\ref{tab:threedimension} presents this multi-dimensional evaluation. The results reveal that a significant portion of generated logs fail on at least one dimension, underscoring the task's complexity. However, the fine-tuned SOLMs consistently produce a higher percentage of fully correct statements (correct position, level, and variables). Qwen2.5-coder-14B-RAG-LoRA again demonstrates its superiority, generating fully correct logs in 27.74\% of cases, the highest of any model. This is closely followed by CodeLlama-13B (25.38\%) and Mistral-7B (25.17\%). Notably, these SOLMs outperform the best proprietary LLM, Deepseek-coder-v3-RAG (24.51\%). This analysis reinforces that SOLMs excel not only on individual metrics but also in producing more integrally correct and practically useful logging statements.}

\begin{figure}
    \centering
    \includegraphics[width=0.9\linewidth]{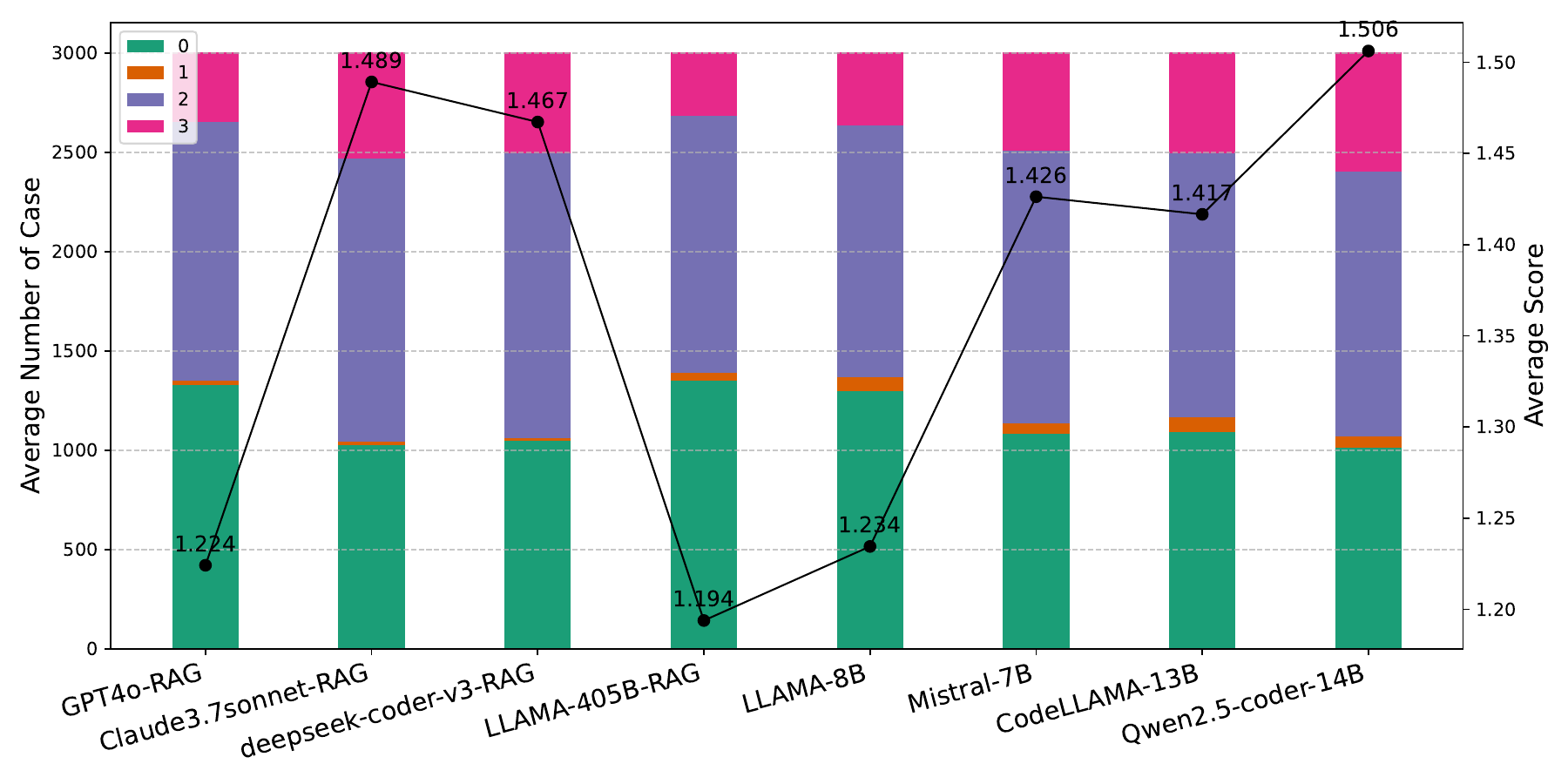}
    \caption{Distribution of LLM Judges Scores for Generated Logging Statement Quality Across Models.}
    \label{fig:res_llmjudger}
\end{figure}

(RQ3.3) Figure~\ref{fig:res_llmjudger} illustrates the score distribution for LLM judges evaluating the quality of logging statements from various methods. First, the bar chart indicates the number of cases receiving each score (0, 1, 2, 3) for different models. Qwen2.5-coder-14B stands out with the highest number of cases scoring 3, alongside the lowest number of cases scoring 0. This distribution suggests that Qwen2.5-coder-14B consistently generates logging statements of higher quality. Second, the trend line representing the average score highlights Qwen2.5-coder-14B achieving the highest average score of 1.506, surpassing all other models, including Claude3.7sonnet-RAG (1.489) and Deepseek-coder-v3-RAG (1.467). 

\begin{tcolorbox}[boxsep=1pt,left=2pt,right=2pt,top=3pt,bottom=2pt,width=\columnwidth,colback=white!95!black,boxrule=0pt, colbacktitle=white!30!black,toptitle=2pt,bottomtitle=1pt,opacitybacktitle=0.4, sharp corners]

\textbf{Findings 8}: Fine-tuned SOLMs outperform both existing methods and LLMs across all evaluated metrics, demonstrating superior logging location accuracy and statement quality. 

\end{tcolorbox}
\subsection{RQ4: How robust are SOLMs in handling diverse real-world scenarios for logging statement generation?}

\textbf{Motivation.}
\added{
In real-world software development, automated logging tools must demonstrate robust performance across diverse challenging scenarios beyond controlled experimental settings. Two critical dimensions of robustness are essential for practical deployment:
}

\added{(RQ4.1) Can SOLMs generalize logging statement generation across diverse code repositories? Software projects exhibit significant heterogeneity in logging practices due to team preferences, domain requirements, and coding conventions. Understanding whether SOLMs can maintain effectiveness when deployed on unseen codebases with different logging practices is crucial for determining their practical viability as general-purpose logging tools.}

\added{(RQ4.2) How does code length affect SOLM performance in automated logging? Production codebases contain methods ranging from simple functions to complex implementations spanning hundreds of lines. Evaluating performance scaling with code length is essential to validate whether our findings about SOLM effectiveness hold across the full spectrum of code complexity.}

\textbf{Approach.}
(RQ4.1) To evaluate the cross-repository generalization ability of SOLMs, we design an experiment that trains models on a subset of repositories from the dataset and tests them on a distinct, non-overlapping set of repositories. We randomly partition the dataset into two groups, each containing five repositories, ensuring diversity in project domains. In each experimental run, we use one group of five repositories as the training set, three repositories from the other group as the validation set, and the remaining two repositories from the same group as the test set. The validation set primarily served to support RAG by providing a pool of examples from which the most similar code snippets are retrieved using the BM25 algorithm. This partitioning ensures that the test set represents unseen projects with potentially different coding styles and logging conventions, simulating real-world cross-repository scenarios. 

The experimental setup involves training on a subset of repositories (R1, R3, R5, R7, R9). It also includes validating on another subset (R2, R4, R6) and testing on unseen repositories (R8, R10) in the first configuration. The second configuration, meanwhile, trains on Apache-dominated repositories (R1, R3, R5, R7, R9) and tests on non-Apache repositories.
For this experiment, we select the two best-performing 7B SOLMs from RQ2.1: Mistral and Qwen2.5-coder. We fine-tune these models using the best-performing strategy identified in prior experiments, combining LoRA with RAG.

\added{(RQ4.2) To comprehensively understand how code length impacts automated logging performance, we construct two test sets from the AL-Bench dataset: Test-Short containing code snippets with less than 512 tokens and Test-Long containing snippets exceeding 512 tokens. This division enables us to analyze whether model performance and the effectiveness of different strategies vary with code complexity, as longer code segments typically contain more intricate control flows and multiple logging-worthy events.}

\added{We evaluate all key configurations explored in previous RQs on both test sets, including the four prompting strategies (base, ICL, RAG, CoT) with Qwen2.5Coder-7B models with all PEFT techniques (Prefix Tuning, Prompt Tuning, LoRA, QLoRA). We also include a representative LLM baselines (i.e., Deepseek-coder-v3) for comprehensive comparison.}

\textbf{Results.}
(RQ4.1) Table~\ref{tab:rq4} presents the performance of two fine-tuned SOLMs, Mistral-7B and Qwen2.5-coder-7B, in automated logging across diverse code repositories.  The results demonstrate the generalization capabilities of SOLMs and highlight the impact of similar logging practices on cross-project performance.

\begin{table}[tbp]
\centering
\caption{Generalization Capabilities for SOLMs Across Diverse Code Repositories.}
\label{tab:rq4}
\resizebox{\textwidth}{!}{%
\begin{tabular}{ccc||c||c||cc|cc|cc}
\toprule
Train Data & Valid Data & Test Data & \multicolumn{1}{c||}{Model} & PA & LA & AOD & PMR & F1 & BLEU-4 & ROUGE-L \\ \midrule
\multirow{2}{*}{R1, R3, R5, R7, R9} & \multirow{2}{*}{R2, R4, R6} & \multirow{2}{*}{R8, R10} & Mistral & 52.97 & 69.30 & 90.74 & 44.31 & 55.87 & 18.80 & 40.15 \\
 &  &  & Qwen2.5coder & 55.54 & 71.67 & 91.86 & 44.67 & 57.20 & 20.21 & 42.22\\ \midrule
\multirow{2}{*}{R2, R4, R6, R8, R10} & \multirow{2}{*}{R1, R3, R5} & \multirow{2}{*}{R7, R9} & Mistral & 44.27 & 51.76 & 79.75 & 43.44 &47.18& 17.25 & 38.46 \\
 &  &   & Qwen2.5coder & 50.71 & 56.19 & 81.96 & 45.71  & 53.30 & 18.48 & 40.50
 \\ \bottomrule
\end{tabular}%
}
\end{table}

The result shows that both Mistral-7B and Qwen2.5-coder-7B exhibit robust performance. Specifically, Qwen2.5-coder achieves a PA of 55.54\% and ROUGE-L of 42.22\%, while Mistral-7B achieves comparable results with a PA of 52.97\% and ROUGE-L of 40.15\%. These metrics indicate that both models successfully generate accurate logging statements and identify appropriate logging locations in unseen repositories, even when the test set includes projects with distinct logging conventions. The high AOD (91.86\% for Qwen2.5-coder) and ROUGE-L scores suggest that the generated logs closely align with ground-truth statements in terms of log level and text similarity. This reveals that SOLMs, when fine-tuned with LoRA and RAG, proficiently generalize across various project areas without experiencing a notable decline in performance.

\begin{tcolorbox}[boxsep=1pt,left=2pt,right=2pt,top=3pt,bottom=2pt,width=\columnwidth,colback=white!95!black,boxrule=0pt, colbacktitle=white!30!black,toptitle=2pt,bottomtitle=1pt,opacitybacktitle=0.4, sharp corners]

\textbf{Findings 9}: SOLMs demonstrate strong generalization capabilities in automated logging, maintaining high performance across diverse, unseen repositories.

\end{tcolorbox}

The performance difference between the two configurations in Table~\ref{tab:rq4} highlights the influence of similar logging practices on cross-project generalization. In the first configuration, where both training and test sets include Apache open-source projects, the models achieve significantly higher performance (e.g., Qwen2.5-coder: PA 55.54\%, ROUGE-L 42.22\%) compared to the second configuration, where the training set comprises Apache projects, but the test set includes non-Apache projects (e.g., Qwen2.5-coder: PA 50.71\%, ROUGE-L 40.50\%). The performance drop in the second configuration (e.g., 4.83\% lower PA and 1.72\% lower ROUGE-L for Qwen2.5-coder) suggests that the absence of shared logging conventions, such as those prevalent in Apache projects (e.g., consistent verbosity levels and formatting styles), reduces the models' ability to generate contextually appropriate logs. This may because Apache projects follow standardized logging guidelines, aiding knowledge transfer during fine-tuning, while non-Apache projects may use diverse, project-specific practices that hinder generalization. This disparity underscores that similarity in logging practices between repositories can enhance cross-project performance.

\begin{tcolorbox}[boxsep=1pt,left=2pt,right=2pt,top=3pt,bottom=2pt,width=\columnwidth,colback=white!95!black,boxrule=0pt, colbacktitle=white!30!black,toptitle=2pt,bottomtitle=1pt,opacitybacktitle=0.4, sharp corners]

\textbf{Findings 10}: Fine-tuning SOLMs with data reflecting similar logging practices, such as those prevalent in Apache open-source projects, significantly enhances their cross-project generalization capabilities.

\end{tcolorbox}

\begin{table}[tbp]
\centering
\small
\caption{\added{Performance Comparison of Different Prompt Strategies and PEFT Techniques on Different Length Code Datasets.}}
\label{tab:rq42}
\begin{tabular}{lccccc}
\toprule
Strategy & Dataset & PA & LA & PMR & ROUGE-L \\ \midrule
\rowcolor{grey}\multicolumn{6}{c}{Different Prompting Techniques} \\ \midrule
\multirow{3}{*}{base} & Short & 4.30 & 58.91 & 27.91 & 27.13 \\
 & Long & 0.80 & 87.50 & 16.67 & 29.44 \\
 & $\Delta$ & $\downarrow$3.50 & $\uparrow$28.59 & $\downarrow$11.24 & $\uparrow$2.31 \\ \midrule
\multirow{3}{*}{base+ICL} & Short & 2.03 & 45.90 & 36.07 & 31.21 \\
 & Long & 0.63 & 63.16 & 21.05 & 28.56 \\
 & $\Delta$ & $\downarrow$1.40 & $\uparrow$17.26 & $\downarrow$15.02 & $\downarrow$2.65 \\ \midrule
\multirow{3}{*}{base+RAG} & Short & 3.40 & 59.80 & 45.10 & 15.10 \\
 & Long & 0.27 & 100.00 & 50.00 & 43.75 \\
 & $\Delta$ & $\downarrow$3.13 & $\uparrow$40.20 & $\uparrow$4.90 & $\uparrow$28.65 \\ \midrule
\multirow{3}{*}{base+COT} & Short & 13.25 & 51.01 & 20.35 & 10.58 \\
 & Long & 6.30 & 77.78 & 16.93 & 32.96 \\
 & $\Delta$ & $\downarrow$6.95 & $\uparrow$26.77 & $\downarrow$3.42 & $\uparrow$22.38 \\
\midrule
\rowcolor{grey}\multicolumn{6}{c}{Different PEFT Techniques} \\ \midrule
\multirow{3}{*}{prefix tuning+RAG} & Short & 25.27 & 65.48 & 44.14 & 35.47 \\
 & Long & 11.25 & 72.58 & 45.92 & 39.36 \\
 & $\Delta$ & $\downarrow$14.02 & $\uparrow$7.10 & $\uparrow$1.78 & $\uparrow$3.89 \\ \midrule
\multirow{3}{*}{prompt tuning+RAG} & Short & 30.04 & 67.63 & 46.12 & 38.74 \\
 & Long & 13.86 & 74.24 & 48.55 & 42.30 \\
 & $\Delta$ & $\downarrow$16.18 & $\uparrow$6.61 & $\uparrow$2.43 & $\uparrow$3.56 \\ \midrule
\multirow{3}{*}{LoRA+RAG} & Short & 62.40 & 68.46 & 52.24 & 44.96 \\
 & Long & 50.37 & 82.70 & 56.97 & 55.86 \\
 & $\Delta$ & $\downarrow$12.03 & $\uparrow$14.24 & $\uparrow$4.73 & $\uparrow$10.90 \\ \midrule
\multirow{3}{*}{QLora+RAG} & Short & 60.21 & 67.87 & 51.55 & 44.03 \\
 & Long & 46.17 & 82.55 & 57.71 & 56.02 \\
 & $\Delta$ & $\downarrow$14.04 & $\uparrow$14.68 & $\uparrow$6.16 & $\uparrow$11.99 \\ \bottomrule
\end{tabular}
\end{table}

\added{(RQ4.2) Table~\ref{tab:rq42} presents a comprehensive analysis of how code length affects SOLM performance across different prompting and PEFT strategies using Qwen2.5Coder-7B.
For prompting techniques without fine-tuning, position accuracy consistently decreases on longer code while logging level accuracy improves substantially. The base strategy shows PA dropping from 4.30\% to 0.80\%, yet LA increases from 58.91\% to 87.50\%. RAG exhibits the most extreme pattern with PA plummeting to 0.27\% but achieving perfect LA of 100\% on long code. COT demonstrates the most balanced degradation with PA declining from 13.25\% to 6.30\% while LA improves from 51.01\% to 77.78\%. This suggests that longer code provides richer semantic context for level determination but severely complicates precise statement positioning.}

\added{PEFT techniques show superior resilience to code length increases. LoRA combined with RAG maintains the strongest position accuracy on long code at 50.37\% while achieving substantial LA improvement from 68.46\% to 82.70\%. QLoRA demonstrates comparable robustness with 46.17\% PA on long code and LA increasing from 67.87\% to 82.55\%. The lighter fine-tuning approaches, prefix and prompt tuning, show moderate resilience with PA values of 11.25\% and 13.86\% respectively on long code, significantly outperforming prompting-only approaches.}

\added{Counterintuitively, variable matching performance and text quality consistently improve on longer code across all configurations. Most strategies achieve better PMR on long code, with LoRA showing improvement from 52.24\% to 56.97\%, indicating that extended context helps models make more informed variable selection decisions. Similarly, ROUGE-L scores universally increase, with LoRA improving from 44.96\% to 55.86\%, demonstrating that additional contextual information enhances text generation quality despite positioning challenges.}

\begin{tcolorbox}[boxsep=1pt,left=2pt,right=2pt,top=3pt,bottom=2pt,width=\columnwidth,colback=white!95!black,boxrule=0pt, colbacktitle=white!30!black,toptitle=2pt,bottomtitle=1pt,opacitybacktitle=0.4, sharp corners]

\added{\textbf{Findings 11}: Code length creates contrasting effects: position accuracy decreases while logging level accuracy, variable matching, and text quality improve, indicating that longer code provides richer semantic context but increases positioning complexity.}

\end{tcolorbox}

\added{Across all configurations, PEFT techniques demonstrate superior resilience compared to prompting-only approaches. Notably, the performance hierarchy established in our earlier experiments remains consistent on longer code: LoRA+RAG maintains the highest position accuracy at 50.37\%. This consistent ranking across different code lengths reinforces the effectiveness of our identified optimal strategies, with LoRA's superior parameter efficiency proving particularly valuable when processing complex, lengthy code segments.}

\begin{tcolorbox}[boxsep=1pt,left=2pt,right=2pt,top=3pt,bottom=2pt,width=\columnwidth,colback=white!95!black,boxrule=0pt, colbacktitle=white!30!black,toptitle=2pt,bottomtitle=1pt,opacitybacktitle=0.4, sharp corners]

\added{\textbf{Findings 12}: The relative performance hierarchy of different strategies remains consistent across code lengths, with LoRA+RAG maintaining its superiority over other PEFT techniques.}

\end{tcolorbox}

\begin{table}[tbp]
\centering
\caption{\added{Performance Comparison of Representative SOLM and LLM on Different Length Code Datasets.}}
\label{tab:rq42_sec}
\begin{tabular}{lccccc}
\toprule
Model & Dataset & PA & LA & PMR & ROUGE-L \\
\midrule
\multirow{3}{*}{Deepseek-coder-v3-RAG} & Short & 65.20 & 68.39 & 49.28 & 41.31 \\
 & Long & 55.23 & 75.98 & 50.57 & 46.88 \\
 & $\Delta$ & $\downarrow$9.97 & $\uparrow$7.59 & $\uparrow$1.29 & $\uparrow$5.57 \\
\midrule
\multirow{3}{*}{Qwen2.5-coder-14B-RAG-LoRA} & Short & 66.20 & 69.92 & 54.53 & 46.51 \\
 & Long & 52.77 & 83.01 & 60.73 & 58.69 \\
 & $\Delta$ & $\downarrow$13.43 & $\uparrow$13.09 & $\uparrow$6.20 & $\uparrow$12.18 \\
\bottomrule
\end{tabular}
\end{table}

\added{Table~\ref{tab:rq42_sec} shows a direct comparison between the best fine-tuned SOLM and the leading LLM baseline. Notably, the fine-tuned SOLM, Qwen2.5-coder, experiences a more pronounced drop in position accuracy ($\downarrow$13.43) than the LLM baseline ($\downarrow$9.97). This suggests that the SOLM's smaller model scale may face greater challenges in pinpointing the exact insertion location within the expanded and more complex context of long code.}

\added{Conversely, this comparison highlights the distinct effectiveness of fine-tuning for content generation. The SOLM outperforms the LLM in its ability to capitalize on the richer context to improve statement quality. For instance, Qwen2.5-coder achieves more improvement in its ROUGE-L score. ($\uparrow$12.18 vs. $\uparrow$5.57) and demonstrates similarly superior gains in both LA and PMR. This indicates that while model scale might influence long-range positional awareness, task-specific fine-tuning is highly effective at teaching the model what to log, enabling it to generate more accurate and contextually relevant content when more information is available.}

\begin{tcolorbox}[boxsep=1pt,left=2pt,right=2pt,top=3pt,bottom=2pt,width=\columnwidth,colback=white!95!black,boxrule=0pt, colbacktitle=white!30!black,toptitle=2pt,bottomtitle=1pt,opacitybacktitle=0.4, sharp corners]

\added{\textbf{Findings 13}: While fine-tuned SOLMs show a sharper decline in position accuracy on long code than LLMs, potentially due to their smaller model scale, they achieve substantially greater improvements in content generation quality. This highlights that fine-tuning effectively trains SOLMs to leverage rich context for generating logging statements.}

\end{tcolorbox}

\section{Discussion}

\subsection{In-depth Analysis: Case Study of Success and Failure}

\subsubsection{Successful Case}

\begin{figure}
    \centering
    \includegraphics[width=1\linewidth]{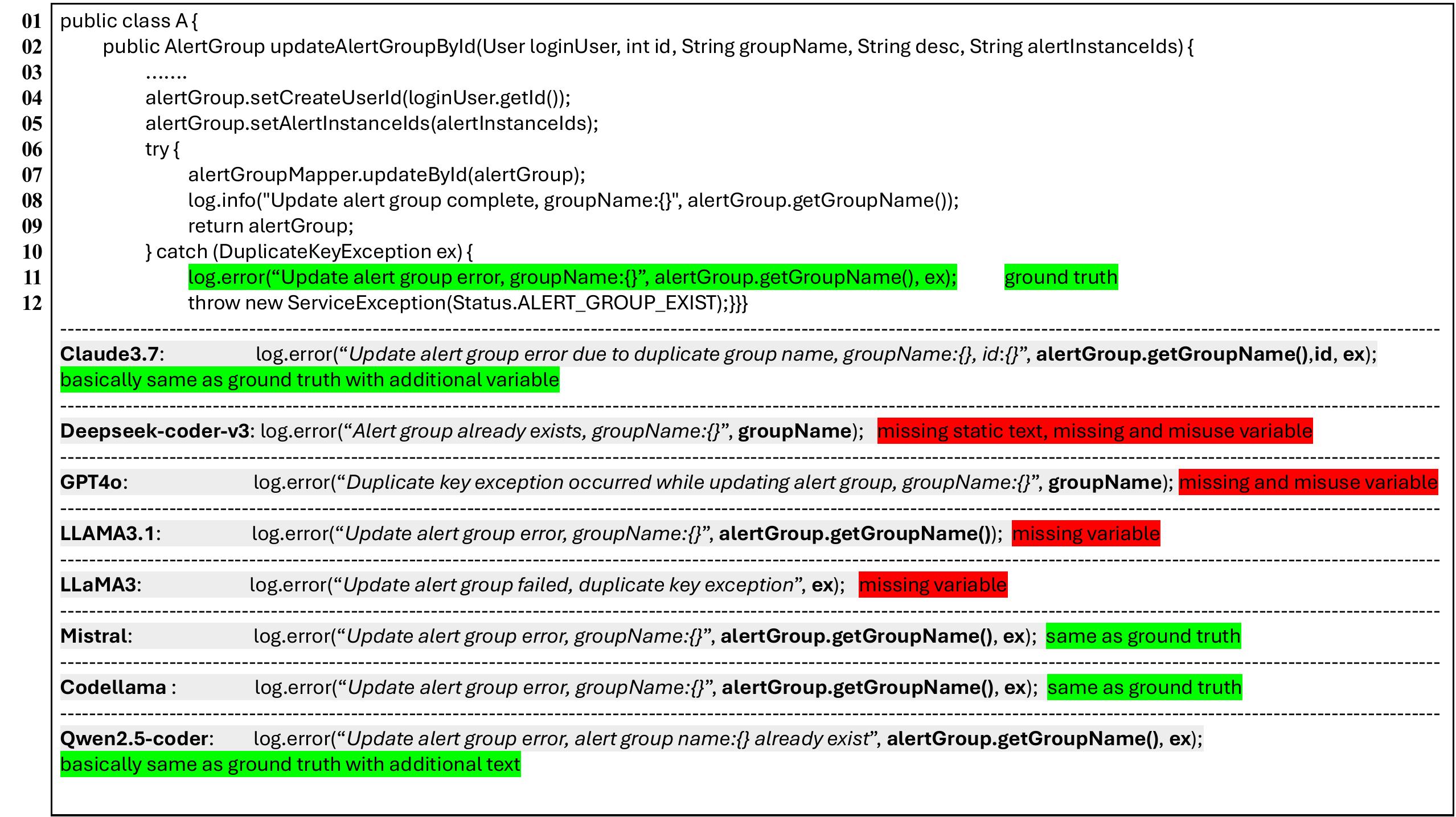}
    \caption{A Case of Generated Logging Statement from Multiple Models.}
    \label{fig:case}
\end{figure}

\added{Figure~\ref{fig:case} presents a case study comparing logging statements generated by different models for an error logging scenario in a Java method handling a DuplicateKeyException. The ground truth logging statement is \texttt{log.error("Update alert group error, groupName:\{\}"), alertGroup.getGroupName(), ex)}. Among the LLMs, Claude3.7sonnet-RAG generates a statement that closely resembles the ground truth but includes an additional variable (\texttt{id}), which may add unnecessary verbosity. Deepseek-coder-v3-RAG and GPT4o-RAG produce statements with missing or misused variables (e.g., using groupName directly instead of alertGroup.getGroupName()), reducing their contextual accuracy. LLaMA3.1-405B-RAG omits the exception variable (ex), limiting its diagnostic utility. In contrast, fine-tuned SOLMs demonstrate superior performance: Mistral-7B-RAG-LoRA and CodeLlama-13B-RAG-LoRA generate statements identical to the ground truth, ensuring both accuracy and relevance. Qwen2.5-coder-14B-RAG-LoRA adds minor additional text ("\texttt{already exist}") but retains all critical components, closely aligning with the ground truth.} 

\subsubsection{Failure Case}

\begin{figure}
    \centering
    \includegraphics[width=1\linewidth]{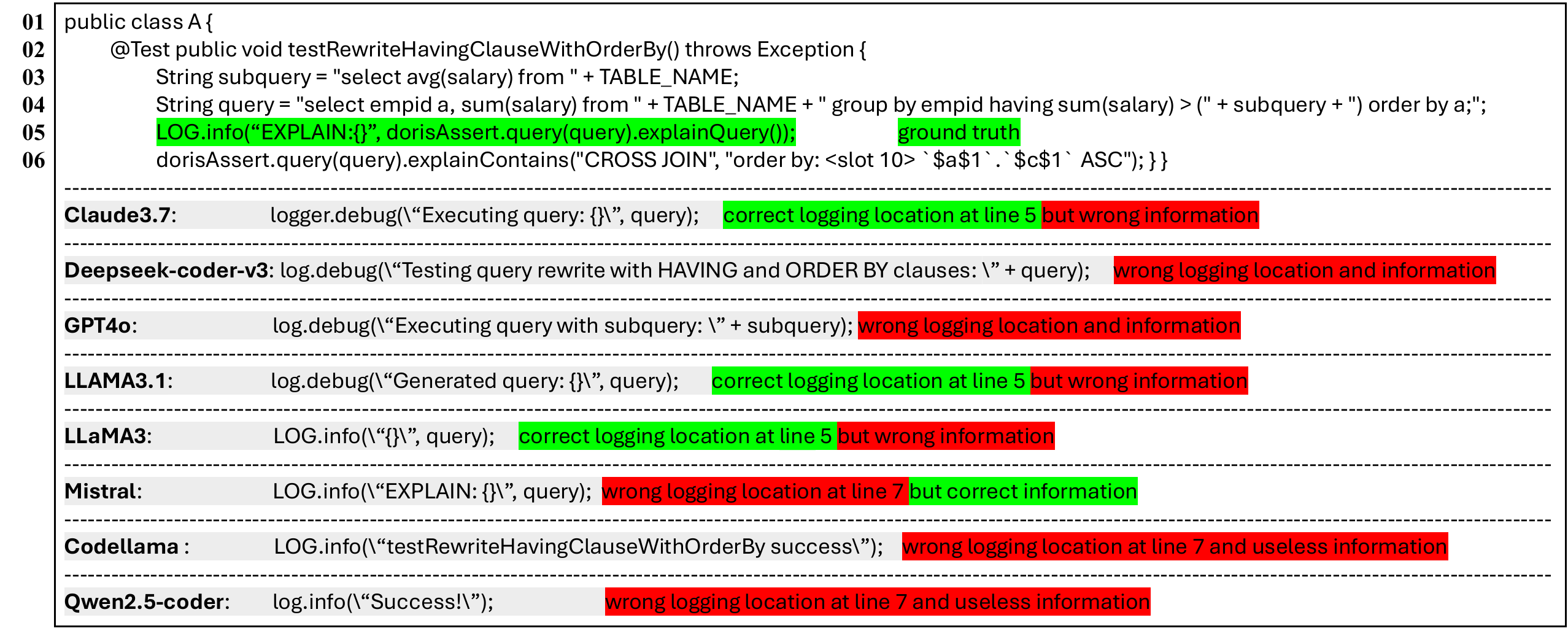}
    \caption{\added{A Case Study on the Failure to Invoke Out-of-Context Methods in Log Generation.}}
    \label{fig:case2}
\end{figure}

\added{Figure~\ref{fig:case2} presents a failure case study comparing logging statements generated by different models. This case study reveals a critical limitation of the models: an inability to generate logging statements that invoke methods not explicitly visible within the immediate code snippet. The diagnostic purpose of the ground-truth log is to capture the output of the \texttt{.explainQuery()} method, which is chained to the \texttt{dorisAssert.query(query)} call. However, most models failed to reason about or infer the existence of this critical but locally unseen method. Instead, they adopted safer, more conservative strategies based only on visible artifacts. For instance, several models correctly identified the logging location but defaulted to logging the simple query variable, demonstrating a failure to explore the object's potential API. More severely, models like Qwen2.5-coder abandoned contextual analysis altogether, generating a generic and uninformative `Success!' message. This tendency to avoid invoking out-of-context methods severely limits their ability to create truly insightful, diagnostic logs and underscores the necessity of fine-tuning on domain-specific API usages to overcome this limitation. Future work could address this limitation by enriching the model's input with structural code representations (e.g., call graphs from static analysis) or by designing more sophisticated, type-aware retrieval mechanisms for RAG that can fetch relevant API usage patterns on demand.}

\begin{figure}
    \centering
    \includegraphics[width=1\linewidth]{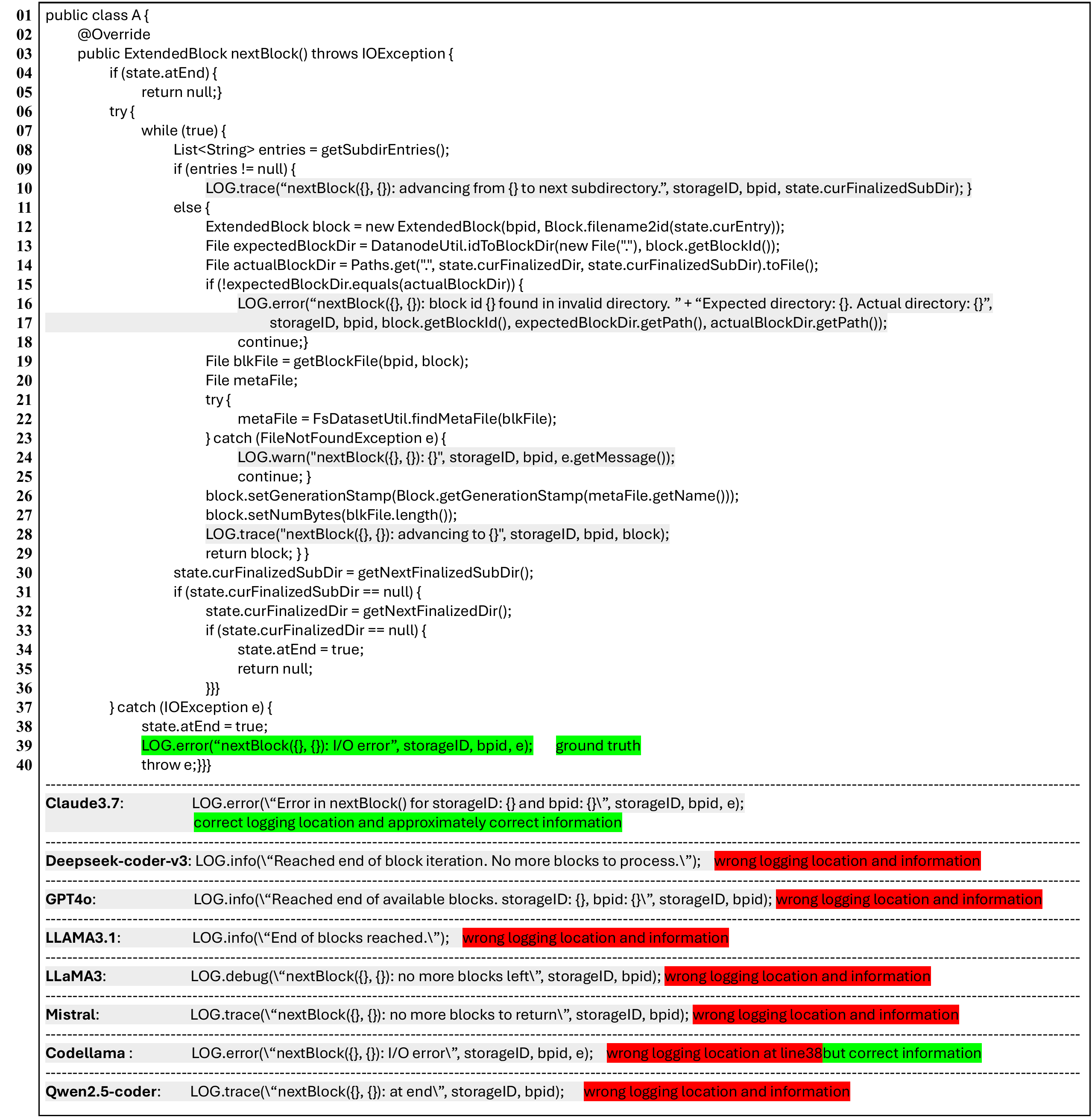}
    \caption{\added{A Case Study on Distraction by Normal Execution Flow in Complex Methods during Log Generation.}}
    \label{fig:case3}
\end{figure}

\added{Figure~\ref{fig:case3} illustrates a critical failure mode in navigating long, complex methods with high logging density. The ground-truth log is situated within a \texttt{catch (IOException e)} block, making its purpose unequivocally tied to error handling. However, the method's length and the presence of five other log statements appear to have primed most models for a different task. Distracted by this high density of existing logs within the try block, models like Qwen2.5-coder seem to have lowered their activation threshold for what constitutes a logging event, becoming overly willing to log the incremental progress of the loop. Consequently, they ignored the distinct exceptional path and prematurely generated logs related to the loop's normal termination, with messages such as \texttt{`Reached end of block iteration'}. This failure likely points to a fundamental limitation of attention-based models, which excel at identifying patterns within the provided context but struggle with reasoning about or `imagining' external APIs not explicitly mentioned. It suggests that without specialized fine-tuning on a project's entire API surface, SOLMs may default to a `what you see is what you get' behavior, limiting their ability to generate truly insightful, diagnostic code.}

\added{The failure to prioritize the exceptional path is further highlighted by the contrasting success of models like Claude3.7, which correctly identified the catch block despite the noisy, log-dense environment of the main logic. The inability of most models to navigate this complexity represents a severe limitation for a practical automated logging tool, as a primary function of logging is to capture unexpected errors. This case suggests that models can be desensitized by existing logs, leading them to misjudge the relative importance of the normal paths versus critical error-handling blocks. Future work should therefore focus on improving models' structural awareness and their ability to reason about control flow priority in lengthy, real-world methods that already contain logging. This case highlights a potential `attention bias' in log-dense code. The model, when exposed to numerous logging examples for normal execution paths, appears to lower its threshold for what constitutes a log-worthy event. It consequently fails to assign a higher priority to the semantically critical but less frequent exceptional path in the catch block. This suggests that future research should focus not just on code semantics, but on teaching models to reason about control flow priority and the relative importance of different execution paths.}

\subsection{Implications and Advice}

\added{Our empirical analysis showcases the potential of fine-tuned SOLMs. Based on our findings, we identify three key implications and provide actionable advice to guide future research and practice in automated logging statement generation.}

\added{}

\added{\textbf{Adopt a Synergistic Optimization Blueprint.} Our results consistently demonstrate that the optimal performance of SOLMs is not achieved through any single technique but through a deliberate, multi-stage optimization process. We found that while `instruct-tuned' models provide a superior foundation over `base' models, their true potential is only unlocked via task-specific fine-tuning, for which LoRA emerged as the most effective and stable PEFT technique. Furthermore, the performance of a LoRA-tuned model is maximally amplified when coupled with a RAG pipeline, which provides critical instance-specific context at inference time. The synergy between task-level specialization (from LoRA) and instance-level specialization (from RAG) consistently yields the best results. This suggests a clear blueprint for developing high-performance, customized code generation tools.}

\begin{tcolorbox}[boxsep=1pt,left=2pt,right=2pt,top=3pt,bottom=2pt,width=\columnwidth,colback=white!95!black,boxrule=0pt, colbacktitle=white!30!black,toptitle=2pt,bottomtitle=1pt,opacitybacktitle=0.4, sharp corners]

\added{\textbf{Implication 1.} Future research and development of SOLM-based tools should adopt a holistic optimization strategy.}

\end{tcolorbox}

\added{\textbf{Re-evaluate the Performance-vs-Resource Paradigm.} A central finding of our work challenges the prevailing "larger is better" paradigm. We demonstrate that a well-optimized, sub-14B parameter SOLM can significantly outperform massive, proprietary LLMs and existing SOTA tools in both logging location and statement quality. This indicates that for specialized SE tasks, targeted adaptation is more critical than raw model scale. However, this does not imply that the smallest models are sufficient; our findings reveal a performance `sweet spot', with models above 3B parameters showing more consistent and meaningful performance gains. This trade-off is further nuanced in complex scenarios; on long code snippets, fine-tuned SOLMs show a greater decline in positioning accuracy than LLMs but achieve substantially larger improvements in content quality, proving more adept at leveraging rich context.}

\begin{tcolorbox}[boxsep=1pt,left=2pt,right=2pt,top=3pt,bottom=2pt,width=\columnwidth,colback=white!95!black,boxrule=0pt, colbacktitle=white!30!black,toptitle=2pt,bottomtitle=1pt,opacitybacktitle=0.4, sharp corners]

\added{\textbf{Implication 2.} Practitioners should shift from defaulting to the largest available models and instead identify a cost-effective `sweet spot' for SOLM size.}

\end{tcolorbox}

\added{\textbf{Prioritize Data Strategy and User Experience for Robustness.} Our study instills confidence in the real-world deployability of SOLMs, which exhibit strong generalization to unseen repositories. However, we found that this robustness is not inherent but is significantly influenced by data and context. The generalization capability of a model is greatly enhanced when it is fine-tuned on a corpus with consistent logging standards, such as those found across Apache Foundation projects. Furthermore, practitioners must be cognizant of the performance trade-offs introduced by code length, where positioning accuracy degrades while content quality improves. The stability of the LoRA+RAG strategy across these varied conditions underscores its reliability.}

\begin{tcolorbox}[boxsep=1pt,left=2pt,right=2pt,top=3pt,bottom=2pt,width=\columnwidth,colback=white!95!black,boxrule=0pt, colbacktitle=white!30!black,toptitle=2pt,bottomtitle=1pt,opacitybacktitle=0.4, sharp corners]

\added{\textbf{Implication 3.} To build robust and generalizable tools, practitioners should prioritize the curation of fine-tuning datasets that reflect consistent coding standards.}

\end{tcolorbox}

\subsection{Analysis of SOLMs’ Advantages}

\subsubsection{Efficiency and Cost-Effectiveness}
\added{From a practical enterprise standpoint, SOLMs present a highly efficient and cost-effective solution for automated logging compared to strategies centered on LLMs. An organization seeking a customized logging tool faces a trade-off: using a commercial LLM via its API involves continuous, per-token operational costs that can become substantial at scale, alongside significant privacy risks from sending proprietary code to a third party. Alternatively, self-hosting and fine-tuning a massive open-source LLM requires a prohibitive investment in computational resources. Our approach offers a compelling middle ground. As demonstrated in our experiments, an SOLM like Qwen2.5-coder-14B can be fine-tuned on enterprise-specific data in under 6 hours on a single A100 GPU to achieve state-of-the-art results (Table 5). This makes the initial customization cost manageable and, more importantly, enables local deployment for inference. Such deployment drastically lowers the long-term cost and energy consumption while completely preserving data privacy, offering a practical and sustainable path for adopting automated logging tools.}

\subsubsection{Privacy and Security}
Privacy preservation is a standout advantage of SOLMs. Li et al.~\cite{li2023exploring} highlight that LLMs often rely on cloud-based APIs, posing risks of proprietary code leakage. In contrast, SOLMs' smaller size enables local deployment, ensuring sensitive code remains secure. This is particularly valuable for companies where strict data protection is paramount, offering a safer alternative for logging generation.

\subsubsection{Adaptability to Enterprise-specific styles}
One of the key challenges in automated logging is adapting to enterprise-specific logging styles and conventions, which vary significantly across organizations due to differences in verbosity levels, error prioritization, or compliance-driven formatting. Our findings demonstrate that SOLMs, when fine-tuned with techniques such as LoRA and RAG, can effectively align with project-specific logging practices. This adaptability is particularly valuable in real-world scenarios where organizations maintain proprietary logging guidelines. Unlike general-purpose LLMs, which struggle to adapt without extensive retraining, SOLMs can be fine-tuned efficiently on internal codebases, ensuring alignment with organizational standards while minimizing computational overhead.

\subsection{Future Work Directions}


\subsubsection{Integration into Development Tools}
To maximize the practical impact of SOLMs in automated logging, future work should focus on integrating these models into widely used development tools, such as integrated development environments (IDEs), and CI/CD platforms. Real-time logging statement suggestions during code authoring or automated insertion during code reviews could streamline the development process and reduce manual effort. For instance, an IDE plugin leveraging a fine-tuned SOLM could analyze code context on-the-fly, recommend logging points, and suggest high-quality logging statements tailored to the project’s conventions. Such integrations would require optimizing SOLMs for low-latency inference and ensuring compatibility with diverse development workflows. Additionally, incorporating user feedback mechanisms into these tools could enable iterative refinement of generated logs, further aligning them with developer preferences.

\subsubsection{Addressing Dynamic Logging Requirements}
Logging practices often evolve during a project’s lifecycle due to changing requirements, such as new debugging needs or compliance regulations. SOLMs must be capable of adapting to these dynamic requirements without requiring extensive retraining. Future research could investigate continual learning techniques to enable SOLMs to incrementally adapt to new logging conventions or project-specific requirements. For instance, online fine-tuning approaches could allow SOLMs to learn from newly added logging statements in a repository, ensuring sustained alignment with evolving practices. Additionally, exploring active learning strategies, where SOLMs query developers for feedback on ambiguous logging scenarios, could further enhance their adaptability.

\subsubsection{Enhancing Long-Context Reasoning for Accurate Positioning}
\added{Our empirical results revealed a key trade-off: while fine-tuned SOLMs excel at leveraging the richer context of long code to improve logging content quality, their accuracy in determining the precise logging location degrades more significantly compared to larger LLMs. This highlights an important direction for future research in enhancing the long-context reasoning of SOLMs for precise automated logging tasks. Future work could explore architectural enhancements that integrate explicit code structure representations like Code Property Graphs (CPGs). Successfully addressing this challenge would significantly improve the robustness of SOLMs, further solidifying their role as a scalable and reliable solution for automated logging in large-scale industrial projects where complex methods are common.}
\label{implications}

\section{Threats to Validity}
\label{threats}

\subsection{Internal Validity}
\subsubsection{Selection of hyperparameters for fine-tuning.} A potential threat to internal validity lies in the selection of hyperparameters for fine-tuning the SOLMs (e.g., learning rate, batch size, prefix length for prefix tuning, rank for LoRA). The study utilized recommended hyperparameters from official sources due to their proven effectiveness. However, these hyperparameters may not be optimal for all models or datasets, potentially introducing bias in the performance results. Suboptimal hyperparameter choices could lead to underperformance or overfitting, affecting the observed effectiveness of SOLMs compared to baseline LLMs or existing methods. To mitigate this, we conducted preliminary experiments to validate the chosen hyperparameters on a subset of the AL-Bench dataset, ensuring reasonable performance. Nonetheless, a more exhaustive hyperparameter search may further enhance model performance and reduce this threat.

\subsubsection{The potential data leakage.} 

\added{A significant threat to the internal validity of our study is the possibility that the AL-Bench dataset may have been part of the large-scale corpora used to pre-train the evaluated SOLMs and LLMs. Since AL-Bench consists of popular open-source projects, it is plausible that its code was included in web scrapes from platforms like GitHub. If such data leakage occurred, the models' performance might be artificially inflated, stemming from the recall of memorized examples rather than from genuine learning of the automated logging task. We acknowledge this critical issue and present several lines of evidence from our experiments that mitigate this threat.}

\added{First, we observed a substantial performance gap between non-fine-tuned base models and their fine-tuned counterparts (e.g., Qwen2.5-coder's Position Accuracy surged from 4.30\% to 62.40\% after fine-tuning). This demonstrates that our task-specific fine-tuning, not pre-existing knowledge, is the primary driver of performance. Second, our cross-repository generalization experiment (RQ4.1) showed that models trained on one set of projects perform robustly on entirely new, unseen ones. This ability to generalize to new codebases is inconsistent with a simple RAG-based project-specific memorization hypothesis.
Finally, our analysis on long code snippets (RQ4.2) revealed a nuanced trade-off: richer context made log position more difficult but improved log content quality. This complex behavior suggests a genuine reasoning process where models leverage semantic context, rather than a simple recall mechanism, which would likely have improved all metrics uniformly.}

\added{Collectively, this evidence strongly suggests that our findings are robust and that the observed performance stems from effective, task-specific adaptation rather than data leakage. Nonetheless, to further address this threat in future work, we recommend evaluating models on proprietary or newly created datasets that are guaranteed to be absent from pre-training corpora.}

\subsection{External Validity}

\subsubsection{The representativeness of the dataset.} A potential threat to generalizability is that the AL-Bench dataset, comprising 10 open-source projects, may not fully represent logging practices in proprietary codebases. We mitigated this by selecting projects from diverse domains (e.g., task scheduling, messaging systems, IoT platforms), ensuring broad coverage of logging requirements. \added{However, similar to prior work~\cite{li2023exploring}, our study focuses exclusively on Java-based projects, which may limit the generalizability of our findings on SOLM performance to other programming languages. This language-specific focus could restrict insights into how SOLMs perform across diverse language ecosystems, potentially affecting their applicability in non-Java contexts.}

\subsubsection{The selection of SOLM.} Another potential threat to the generalizability of our findings lies in the selection of specific SOLMs evaluated in this study. While these models were chosen based on their established performance in software engineering tasks, they may not fully represent the diversity of available SOLMs or future advancements in model architectures. For instance, other SOLMs with different pre-training datasets, architectural designs (e.g., transformer variants or mixture-of-experts models), or domain-specific optimizations might exhibit varying performance in automated logging. To mitigate this threat, we selected models with broad applicability in code-related tasks and ensured they were fine-tuned using techniques (e.g., LoRA and RAG) to align with logging-specific requirements. However, future work should explore a wider range of SOLMs, including those with different training corpora or specialized architectures, to validate the robustness of our findings across diverse model ecosystems.

\subsection{Construct Validity}
\subsubsection{Adequacy of evaluation metrics for logging quality.} 
\added{A potential threat to construct validity lies in whether the chosen evaluation metrics fully capture the quality of generated logging statements. We identified two key limitations in the traditional metrics used. First, metrics for static text evaluation, such as BLEU-4 and ROUGE-L, are confined to lexical similarity and often overlook semantic consistency. A generated statement might be semantically identical or superior to the ground truth but use different wording, causing it to receive a low score unfairly. Second, the PA metric requires an exact match with the ground-truth location, which is overly rigid. In practice, multiple locations within a code block might be functionally equivalent for logging, but the PA metric would penalize these valid, alternative placements. 
To mitigate these inherent metric-based limitations, we incorporated the 
LLM-as-a-judge approach to holistically assess the overall quality of the generated logging statements. Unlike rigid metrics, the LLM judges evaluate each statement based on a comprehensive scoring guideline that prioritizes semantic relevance, contextual appropriateness, and syntactic correctness. This method allows for a more nuanced assessment; for instance, a statement that is semantically correct but lexically different, or one placed in a functionally equivalent but not identical location, may still receive a high score. This qualitative evaluation, therefore, compensates for the weaknesses of traditional metrics and provides a more practical and robust assessment of a model's true performance.}

\subsubsection{Representativeness of the LLM judger result.} The use of an LLM to evaluate the quality of generated logging statements introduces a construct validity threat if the LLM’s judgments do not align with human developer preferences. While the LLM-as-a-judge approach has shown promise in software engineering tasks~\cite{wang2025can}, its scoring may not fully capture nuanced aspects of log quality, such as clarity, relevance to specific debugging scenarios, or adherence to project-specific logging conventions. Misalignment between LLM and human judgments could lead to over- or underestimation of SOLM performance. Therefore, incorporating human evaluations or domain-specific rubrics in future work could enhance the alignment between the LLM judger and practical logging needs.

\section{Related Work}
\label{relatedwork}

\subsection{Automated Logging}
Traditionally, the automation of logging statements is divided into two primary stages~\cite{chen2021survey,he2021survey}: the identification of logging locations and the creation of logging statements. These stages are respectively denoted as where to log and what to log~\cite{zhong2025LogUpdater}. In addressing the complexities associated with determining where to log, various methodologies have been investigated by researchers to identify appropriate logging locations within source code~\cite{jiaSMARTLOGPlaceError2018,liWhereShallWe2020,yaoLog4PerfSuggestingLogging2018,zhaoLog20FullyAutomated2017,zhuLearningLogHelping2015,yuanBeConservativeEnhancing2012,liStudyingSoftwareLogging2018}. Regarding what to log, the generation of logging statements is usually segmented into three specific subtasks: the generation of logging text~\cite{dingLoGenTextAutomaticallyGenerating2022}, the selection of logging variables~\cite{liuWhichVariablesShould2019,yuanImprovingSoftwareDiagnosability2012}, and the prediction of the logging level~\cite{liWhichLogLevel2017,liDeepLVSuggestingLog2021,liuTeLLLogLevel2022,mizouchiPADLADynamicLog2019}.

The latest methodology offers a solution for the automatic generation of logging statements, addressing the selection of logging locations, determining the levels of statements, composing content, and identifying variables in a single step. Mastropaolo et al.~\cite{mastropaoloUsingDeepLearning2022} introduced LANCE, a pioneering comprehensive tool that creates complete logging statements powered by T5. In addition to this, they developed LEONID~\cite{mastropaolo2024log}, which integrates deep learning with information retrieval techniques to enhance performance. Meanwhile, Xu et al.~\cite{xu2024unilog} presented UniLog, grounded in the in-context learning framework of LLMs. Additionally, Xie et al.~\cite{xie2024fastlog} introduced FastLog, which is capable of swiftly generating and inserting entire logging statements. Furthermore, Li et al.~\cite{li2024go} proposed SCLogger, noted as the first approach to generate contextualized logging statements utilizing inter-method static context.

In this paper, we distinguish our work by focusing on the use of SOLMs for automated logging, addressing the limitations of LLMs in terms of privacy, computational efficiency, and adaptability to enterprise-specific logging practices. Unlike prior proposed approaches, which predominantly rely on LLMs, our study leverages fine-tuned SOLMs. This enables local deployment, mitigating privacy risks associated with cloud-based LLM APIs, and significantly reduces computational overhead, aligning with sustainable computing goals. Furthermore, our comprehensive evaluation using the AL-Bench dataset \citep{tan2025ALBench} demonstrates SOLMs' robust generalization across diverse, unseen repositories, a critical aspect not extensively explored in prior work. By systematically investigating prompting strategies and fine-tuning techniques, we provide a scalable and practical solution for automated logging that balances performance with resource constraints, offering a viable alternative for real-world software development.

\subsection{Studies in Logging Practices}
Advancements in logging within software engineering have sparked a growing interest in exploring logging practices across various domains.
Zeng et al.~\cite{zeng2019Studying} and Chen\cite{chenCharacterizingDetectingAntiPatterns2017} extended the work of Yuan et al.~\cite{yuanCharacterizingLoggingPractices2012} by analyzing log statements in Android and Java systems, revealing the widespread occurrence of logging in these environments. Kabinna et al.~\cite{kabinna2016Examining} investigated how changes such as bug fixes, feature enhancements, and code refactoring often lead to revisions in logging statements. Lai et al.~\cite{lal2015two} provided insights into logging code constructs at both file-level and block-level, addressing nine key research questions focused on statistical and content analysis.
Li et al.~\cite{liQualitativeStudyBenefits2021} conducted a detailed qualitative study of the advantages and challenges associated with logging in software development, while Zhou et al.~\cite{zhou2020mobilogleak} explored the connection between logging practices and data leaks in mobile applications. Zhao et al.~\cite{zhao2023studying} analyzed IDs within log statements, proposing a straightforward approach to inject IDs to reduce information loss and examining the extent of information gained through this technique. Li et al.\cite{li2023did} investigated the characteristics and practical importance of dynamic variables, proposing a variable-aware log abstraction technique.
Li et al.\cite{li2023exploring} introduced a study on LLM-assisted logging statement generation, demonstrating that prompt-based zero-shot or few-shot learning significantly enhances the generalization capabilities of LLMs. Tan et al.~\cite{tan2025ALBench} proposed AL-Bench, which is a comprehensive benchmark designed specifically for automatic logging tools.

\section{Conclusion}

In this paper, we explore solutions to the challenges associated with manual logging statement generation by comprehensively investigating the potential of SOLMs as a more viable alternative to resource-intensive and privacy-concerning LLMs. We conduct an extensive, large-scale empirical study to systematically evaluate the efficacy of four prominent SOLMs. This evaluation encompasses various prompt strategies, such as RAG; PEFT techniques, with a focus on LoRA; the impact of different model sizes; and the comparative performance of base versus instruction-tuned model types.

Our key findings provide several critical insights into leveraging SOLMs for automatic logging. We demonstrate that (1) RAG significantly enhances the performance of SOLMs in generating logging statements, and (2) LoRA proves to be a highly effective PEFT technique, enabling substantial improvements with minimal trainable parameters. (3) While larger SOLMs generally yield better results, this is balanced by increased computational demands, highlighting an important performance-resource trade-off. Furthermore, (4) instruction-tuned SOLMs consistently outperform their base counterparts, benefiting from their inherent instruction-following capabilities. Most notably, (5) our research establishes that fine-tuned SOLMs, particularly Qwen2.5-coder-14B, can surpass existing specialized logging tools and even larger LLM baselines in both the accuracy of predicting logging locations and the quality of the generated statements. These findings are further corroborated by LLM-as-a-judge evaluations, which confirm the high quality of SOLM-generated outputs. Additionally, (6) we find that SOLMs exhibit robust generalization capabilities across diverse, unseen code repositories, underscoring their practical applicability.

In conclusion, this study provides strong evidence that appropriately fine-tuned SOLMs offer a powerful, efficient, privacy-preserving, and adaptable solution for automated logging. By making our methodology, datasets, and results publicly available~\cite{artifacts}, we aim to stimulate further research and development in this domain, ultimately contributing to improved software maintenance practices.

\section*{Acknowledgement}
The work described in this paper was supported by the Research Grants Council of the Hong Kong Special Administrative Region, China (No. CUHK 14206921 of the General Research Fund).
\bibliographystyle{ACM-Reference-Format}
\bibliography{bibfile}


\end{document}